\documentclass[journal=nanoletters,manuscript=letter]{achemso}

\usepackage{chemformula} % Formula subscripts using \ch{}
\usepackage[T1]{fontenc} % Use modern font encodings

\author{Elissa Klopfer}
\affiliation{Department of Materials Science and Engineering, Stanford University, Stanford, California 94305, USA}
\altaffiliation{These authors contributed equally to this work}
\email{eklopfer@stanford.edu}
\author{Sahil Dagli}
\affiliation{Department of Materials Science and Engineering, Stanford University, Stanford, California 94305, USA}
\altaffiliation{These authors contributed equally to this work}
\author{David Barton III}
\affiliation{John A. Paulson School of Engineering and Applied Sciences, Harvard University, Cambridge, Massachusetts 02139, USA}
\author{Mark Lawrence}
\affiliation{Department of Electrical and Systems Engineering, Washington University in St. Louis, St. Louis, Missouri 63130, USA}
\email{markl@wustl.edu}
\author{Jennifer A. Dionne}
\affiliation{Department of Materials Science and Engineering, Stanford University, Stanford, California 94305, USA}
\alsoaffiliation{Department of Radiology, Stanford University, Stanford, California 94305, USA}
\email{jdionne@stanford.edu}

\title[]
  {High quality factor silicon-on-lithium niobate metasurfaces for electro-optically reconfigurable wavefront shaping}

\begin{document}

\begin{abstract}
Dynamically reconfigurable metasurfaces promise compact and lightweight spatial light modulation for many applications, including LiDAR, AR/VR, and LiFi systems. Here, we design and computationally investigate high quality factor silicon-on-lithium niobate metasurfaces with electrically-driven, independent control of its constituent nanobars for full phase tunability with high tuning efficiency. Free-space light couples to guided modes within each nanobar via periodic perturbations, generating quality factors exceeding 30,000, while maintaining bar spacing <$\lambda$/1.5. We achieve nearly 2$\pi$ phase variation with an applied bias not exceeding $\pm$ 25 V, maintaining reflection efficiency above 91\%. Using full-field simulations, we demonstrate a high angle, 51{\textdegree}, switchable beamsplitter with a diffracted efficiency of 93\%, and an angle-tunable beamsteerer, spanning 18-31{\textdegree}, with up to 86\% efficiency, all using the same metasurface device. Our platform provides a foundation for highly efficient wavefront shaping devices with a wide dynamic tuning range capable of generating nearly any transfer function.
\end{abstract}

\subsubsection{Keywords}
reconfigurable metasurface, electro-optics, high-Q, multifunctionality, beamsteering

\section{}
The ability to deterministically shape and control wavefronts is essential for optical technologies spanning communication, computation, and sensing\cite{Savage2009}. Lightweight, compact, and mobile platforms are especially important, and in the past decade, have been accelerated by advances in metasurfaces. These ultra-thin surfaces are composed of subwavelength antennas that precisely control the phase, polarization, and amplitude of transmitted or reflected light.  Utilizing geometric patterning, the optical response of metasurfaces can be tailored to realize beamsteering\cite{Yu2011e,Wang2018k}, lensing\cite{Chen2019c,Khorasaninejad2015c,Hu2016}, and holography\cite{Huang2013b}, among other transfer functions, each in a subwavelength-thick platform with comparable performance to bulk optics. 

While traditional passive metasurfaces are limited to performing a specific application pre-determined by their architecture, reconfigurable metasurfaces dynamically change their optical wavefront\cite{Shaltout2019f}. A variety of modulation techniques have been explored, including electro-optic, thermo-optic, mechano-optic, and nonlinear effects in materials\cite{She2018a,Wang2016b,Howes2020b,Wang2020a,Shalaginov2021a,Zhao2018,Li2018}. Among these, electrical tuning is the only reasonable approach for incorporation into near-term commercial devices. To date, modulation with electro-optic effects has been shown using liquid crystals\cite{Li2019s}, MEMS\cite{Holsteen2019b}, the electro-optic Stark effect\cite{Wu2019d}, epsilon-near-zero materials\cite{Shirmanesh2020d}, or tuned carrier concentration\cite{Park2020e}, and has reached switching speeds exceeding the kHz range\cite{Park2020e,Wu2019d}. Of particular interest is the ability to individually address metasurface elements; recent advances here have resulted in reconfigurable wavefront shaping devices capable of multiple transfer functions, akin to spatial-light-modulators, including tunable beamsteering and lensing\cite{Li2019s,Wu2019d,Shirmanesh2020d,Park2020e}. However, to date, metasurface spatial-light-modulators have suffered from a combination of low directivity, which quantifies the contribution from spurious diffraction artifacts, low diffraction efficiency and a small field of view\cite{Kim2021}. For example, while a field of view as large as 80{\textdegree} has been measured with reflective plasmon-based schemes, the reliance on strong absorption resonances to amplify the index modulation has kept diffraction efficiencies below 10\%\cite{Shirmanesh2020d,Huang2016d}. The limited range of accessible phase levels as well as unwanted amplitude changes accompanying the intended phase settings has also resulted in low directivity. Higher diffraction efficiencies, although still less than 50\%, have been observed with a transmissive approach using liquid-crystal, yet the low spatial resolution of the constituent elements in this case again resulted in poor directivity and limited the field of view to less than 22{\textdegree}\cite{Li2019s}. 

Here, we propose a fully reconfigurable silicon-on-lithium niobate metasurface capable of high overall efficiency and high accuracy wavefront shaping, with a large dynamic tuning range. Our design leverages high quality factor (high-Q) nanobars to arbitrarily control the phase response of each metasurface element, which we individually address through transparent conducting oxide contacts on each bar. With the ability to individually address metasurface elements, we can construct phase gradient transfer functions defined by the applied field, rather than varying the size, spacing, or geometry of the constituent nanobars. By changing the nanobar biasing configuration, we show a single metasurface can be tuned to steer light to different angles with high diffraction efficiency. We further demonstrate how our metasurface can switch between multiple transfer functions through coupled electrostatic and electromagnetic full field finite element simulations. Specifically, we show how an applied field can modulate the metasurface to act as beamsplitting or beamsteering devices. In beamsteering, we demonstrate reconfigurability up to 51{\textdegree}, corresponding to a 102{\textdegree} field of view, with an efficiency of 93\%. As a beam-steerer, we show how biasing can result in diffracted  angles spanning 18-31{\textdegree} with efficiencies as high as 86\%. Our work is the first to show a high-Q metasurface design to access a phase-voltage library for near-limitless phase gradient metasurfaces.

%\section{Design of electro-optically tunable high-Q metasurface}

Figure 1a illustrates our device design. The metasurface is composed of a series of identical silicon-on-lithium niobate nanobars, which we tune to control the optical phase and resulting reflected wavefront. As seen in Figure 1b, the nanobars are 500 nm wide across the 220-nm-thick Si layer. Beneath the Si is 100-nm-thick x-cut lithium niobate (LNO), designed with a 10{\textdegree} taper angle, commensurate with fabrication constraints\cite{Jiang2019a,Li2020a}. The high index of Si localizes light and serves as a waveguide, while the LNO is used as the active, electro-optic material. The thickness of Si was selected to  enhance the modal overlap in the active LNO layer by promoting otherwise tightly confined modes to leak out of the Si. Similarly, the thickness of LNO was selected to encapsulate the spatial extent of the leaked high-Q mode, while minimizing the distance between the electrostatic contacts. Our electrostatic contacts consist of two 50 nm layers of transparent conducting oxide (TCO), here modeled as indium tin oxide (ITO), on top of the Si and beneath the LNO\cite{Dabirian2017}. Maintaining minimal distance between the electrical contacts raises the resulting electrostatic field intensity experienced by the nanobar. The ITO top contact is the same width as the Si to electrically tune each nanobar individually, while the bottom contact is continuous and acts as the ground.

The center-to-center distance of the nanobars is 1000 nm, chosen to significantly separate the resonators to decrease coupling while also maintaining the subwavelength nature of each metasurface unit cell element.  Two additional nanofins with 100 nm width are included on either side of the nanobars to act as isolators between nanobars to suppress coupling and voltage crosstalk, ensuring the  high-Q  modes operate independently  from  one  another while maintaining subwavelength separation\cite{Jahani2014a,Jahani2018b}. The device operates in reflection with the addition of a metallic layer (in this case gold)\cite{Ordal1987}, beneath an insulating layer of silicon dioxide (SiO$_2$); we position the gold reflector 610 nm below the bottom contact, far enough away to reduce absorption while also maximizing reflection (further criteria for this metal plate depth is discussed in the Supporting Information, see Figure S3). Figure 1b includes the  geometry of the metasurface unit cell, including the nanobar resonator and two nanofin isolators.

The introduction of periodic notches into each nanobar induces a high-Q resonance in the form of a guided mode resonance (GMR). It is well known that Si waveguides support guided modes\cite{Johnson1999c,Wang1993c,Hsu2016d} due to high index contrast.  Applying a periodic perturbation to the waveguide provides additional momentum that allows the guided modes to couple to a normally incident plane wave\cite{Lawrence2018i,Kim2019k}. Such GMRs arise from the imposed Bloch condition on the waveguide dispersion due to discrete translation symmetry (see Figure S1). When the perturbation is kept small, these GMRs are accompanied by highly enhanced near fields and long resonant lifetimes\cite{Lawrence2020d}, and when embedded into metasurfaces enable efficient nonlinear\cite{Lawrence2018i,Lawrence2019b} and dynamic\cite{Klopfer2020c,Barton2020c} diffraction control. We select a notch period of 650 nm to introduce a resonance at approximately 1565 nm (see Figure S1). The inset of Figure 1b shows the symmetric pair of notches, 100 nm wide and 50 nm deep. The size of the notch influences the guided mode resonance radiative coupling strength, with shallower notches resulting in higher Q factors, as shown in Figure 1c (COMSOL Multiphysics).

Figure 1d and e show the optical electric field distribution for the low-Q unnotched and high-Q notched cases of our metasurface design when illuminated by light with an x-polarized electric field incident from the z-direction. The highly resonant high-Q structured metasurface enhances the electric field by up to 280$\times$ in the Si and 83$\times$ in the LNO compared to the non-resonant case. The reflection spectrum of our metasurface is depicted in Figure 1f. For the nanobar structure with no perturbations, we observe generally high reflection greater than 98\% across the wavelength range. When periodic perturbations are included, we observe a resonant dip in the reflection to approximately 91\%, with a Q factor exceeding 33,000. The inclusion of the reflecting plate forms an additional cavity, and the resulting resonance Q factor is a function of both the GMR and the Fabry-Perot behavior of this cavity.

\begin{figure}[htbp]
\centering\includegraphics[width=7cm]{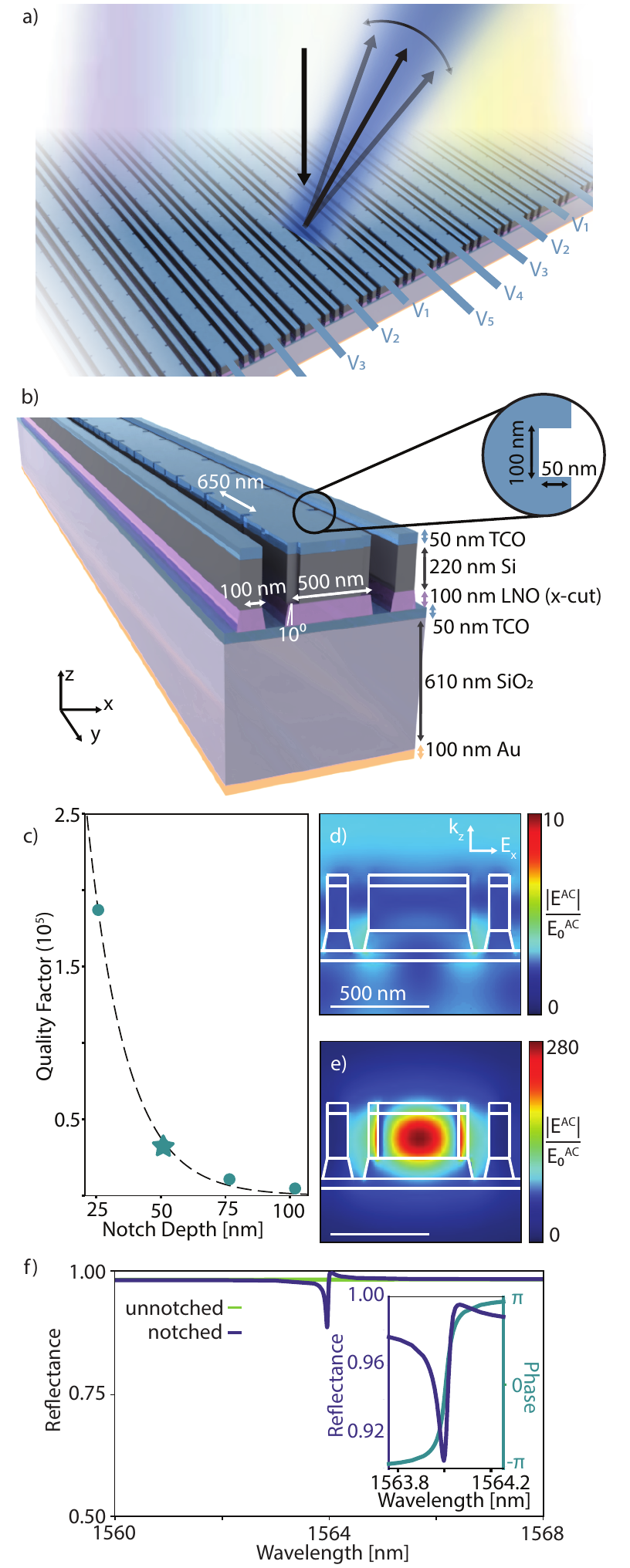}
\caption{High-Q metasurface geometry and spectral characteristics. (a) High-Q metasurface schematic operating as a tunable beamsteerer with applied bias. (b) Geometry of the metasurface unit cell, consisting of a 500 nm wide high-Q nanobar resonator and two 100 nm wide isolator nanofins. (c) Q factor dependence on varying notch depth. (d) Normalized electromagnetic near-field for the nanobar without notches at $\lambda$=1563.96 nm. (e) Normalized electromagnetic near-field for the nanobar with 50$\times$100 nm notches every 650 nm at $\lambda$=1563.96 nm. (f) Spectra of the unnotched (green) and notched (blue) metasurface platform. Inset zooms-in to the resonant feature and phase variation, with a calculated quality factor of 33,253.}
\end{figure}

In addition to the high efficiency reflection, the high quality factor resonance is accompanied by a  2$\pi$ phase variation in the reflected light (as described in the SI, as a transmissive device of the same design exhibits 1$\pi$ of phase variation, occurring near a null in transmission, see Figure S2). Such a full 2$\pi$ phase space allows for nearly any transfer function to be constructed - ranging from simple beamsplitters and beamsteerers to more complex systems such as lenses and holograms\cite{Davoyan2020}. We utilize the electro-optic effect in the LNO layer to individually address each nanobar.  Here, a DC applied field in the appropriate direction modifies the LNO permittivity according to:
\begin{equation}
    \Delta\epsilon_{xx} = -r_{33}n_e^4E_z^{app}
\end{equation}
\begin{equation}
    \Delta\epsilon_{yy} = \Delta\epsilon_{zz} = -r_{13}n_o^4E_z^{app}
\end{equation}
Here, $r_{13}$= 9 pm/V, and $r_{33}$= 31 pm/V are the LNO electro-optic coefficients\cite{Weis1985c}, $n_o$ = 2.21 and is the ordinary refractive index of LNO, and $n_e$ = 2.14 and is the extraordinary refractive index of LNO. $E_z^{app}$ is the electric field applied with the voltage bias across the TCO contacts. We simulate an applied voltage in our metasurface design by calculating the DC electric field from a voltage applied across the TCO layers above a given nanobar, as shown schematically in Figure 2a. The electrostatic DC field resulting from applying 1 V to the top contact is shown in  Figure 2b. The high DC relative permittivity of LNO ($\epsilon_{11,DC}$ = 46.5, $\epsilon_{33,DC}$ = 27.3)\cite{Weis1985c} results in a decreased applied DC electric field strength in the LNO layer, where the electric field is about four times stronger in the Si than the LNO. We compensate for this by minimizing the distance between the TCO contacts to increase the overall electric field. 

Since high-Q  resonances  are  very  sensitive to subtle changes in refractive index, even small changes associated with otherwise weak electro-optic effects tune the device at reasonable voltages\cite{Wang2019e,Chen2013g,Weigel2018d,Witmer2017d,Barton2021d}. Figure 2c shows spectral shifts of the high-Q resonance for $\pm$10 V. As seen, the resonance red-shifts with positive applied bias and blue shifts with negative applied bias. This shift is accompanied by changes to the phase and amplitude of the reflected light at the unbiased resonant wavelength, $\lambda=1563.96$ nm, shown in Figure 2d. Our high-Q resonance changes its spectral position by a full linewidth with biases of $\pm$25 V, allowing us to  achieve full 2$\pi$ phase tunability of individual nanobars with reasonable applied electric fields. We note that the required voltage could be further reduced by increasing Q or increasing the optical and DC field overlap in the LNO. 

\begin{figure}[htbp]
\centering\includegraphics[width=7cm]{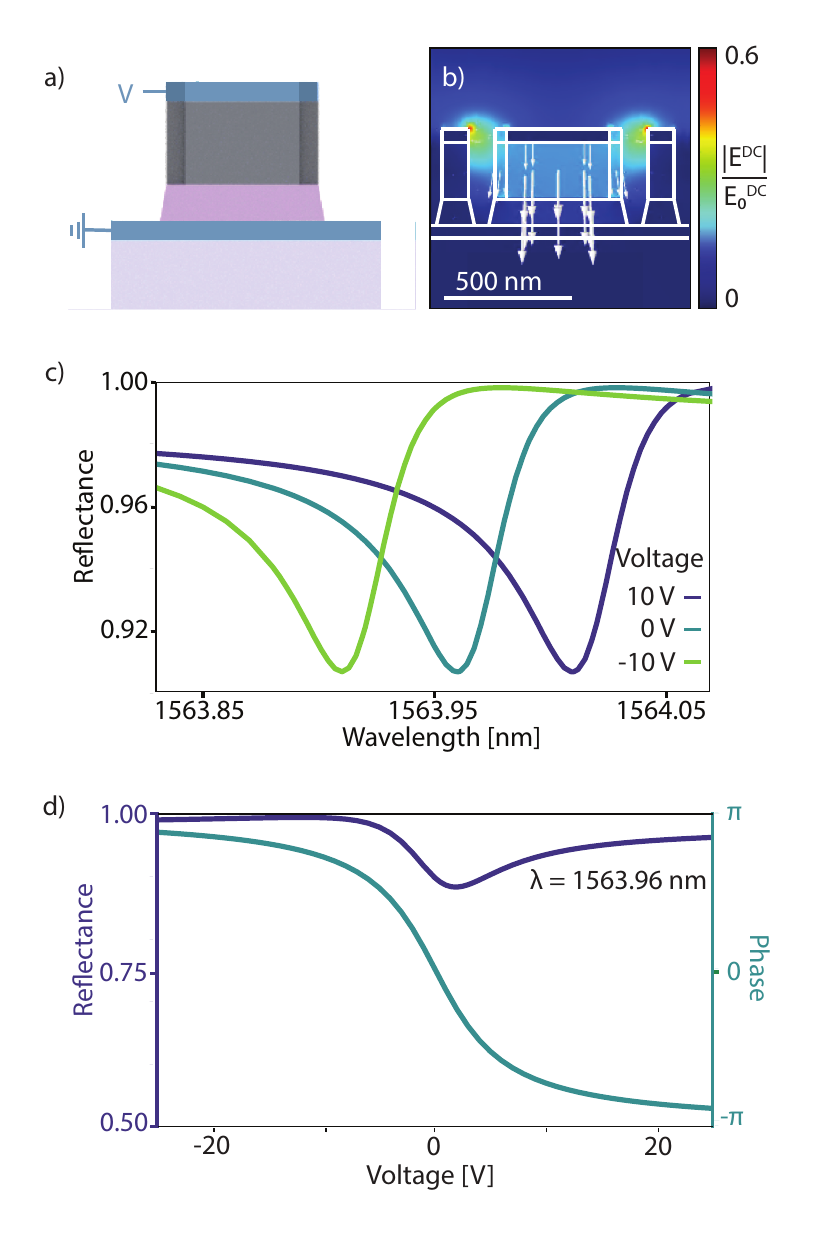}
\caption{Electro-optic shifts of the High-Q resonance. (a) Schematic of an isolated nanobar resonator where voltage can be applied to the transparent conducting oxide (TCO) to induce an electrostatic bias across the resonator. (b) Normalized electrostatic near-field induced by applying 1 V to the layer of TCO above the nanobar resonator with  the continuous layer of TCO acting as ground. (c) Spectra of the high-Q resonance shifting with applied voltages: -10 V (green), 0 V (cyan), +10 V (blue). (d) Reflectance (blue) and associated phase delay (cyan) of the nanobar resonator as a function of applied voltage ($\lambda$=1563.96 nm).}
\end{figure}

%\section{Numerical demonstration of electro-optic beamsplitting and tunable beamsteering}

Figure 3 shows the design of a switchable beamsplitter, which we construct from a pair of nanoantennas resonating $\pi$ out of phase of one another\cite{Lawrence2020d}. By applying alternating biases to every other nanobar, schematically shown as a supercell of 2 nanobars in Figure 3a, we electro-optically shift the phase delay of neighboring nanobars $\pi$ out of phase of each other. Figure 3b shows the resulting phase versus voltage for nanobars without nanofins between them; at voltages of $\pm$21 V, we achieve a phase delay difference of $\pi$ at the resonant wavelength  1562.65 nm. Without nanofins, the application of these voltages to alternating nanobar results in a $\pm$1st combined diffraction efficiency of only  49\%, as seen in Figure 3c and the field plot of Figure 3d. In contrast, when nanofins are included to properly isolate the high-Q modes from one another, significant beamsplitting can be observed (Figure 3e,f) First, we note that the addition of nanofin isolators does not affect the metasurface’s behavior without bias, apart from the resonance being shifted slightly to 1563.96 nm. The inclusion of nanofins changes the phase response with applied voltage (Figure 3e), and we therefore utilize $\pm$11.3 V for our beamsplitting condition, as illustrated in Figure 3e. In this case, the $\pm$1st diffraction order achieves 93\% combined efficiency on resonance at 1563.96 nm (Figure 3f). This higher efficiency corresponds to clear beamsplitting in the field plot in Figure 3g, where light is directed to $\pm$51.4{\textdegree} off normal. Put otherwise, by carefully engineering high-Q resonances, maintaining their isolation with additional nanostructures, and applying particular voltage biases, we are able to freely construct a dynamically switchable beamsplitter. Further discussion on the efficacy of the nanofins, including nearfield plots of the nanobar resonators, is found in the Supporting Information (see Figure S4 and S5).

\begin{figure}[htbp]
\centering\includegraphics[width=7cm]{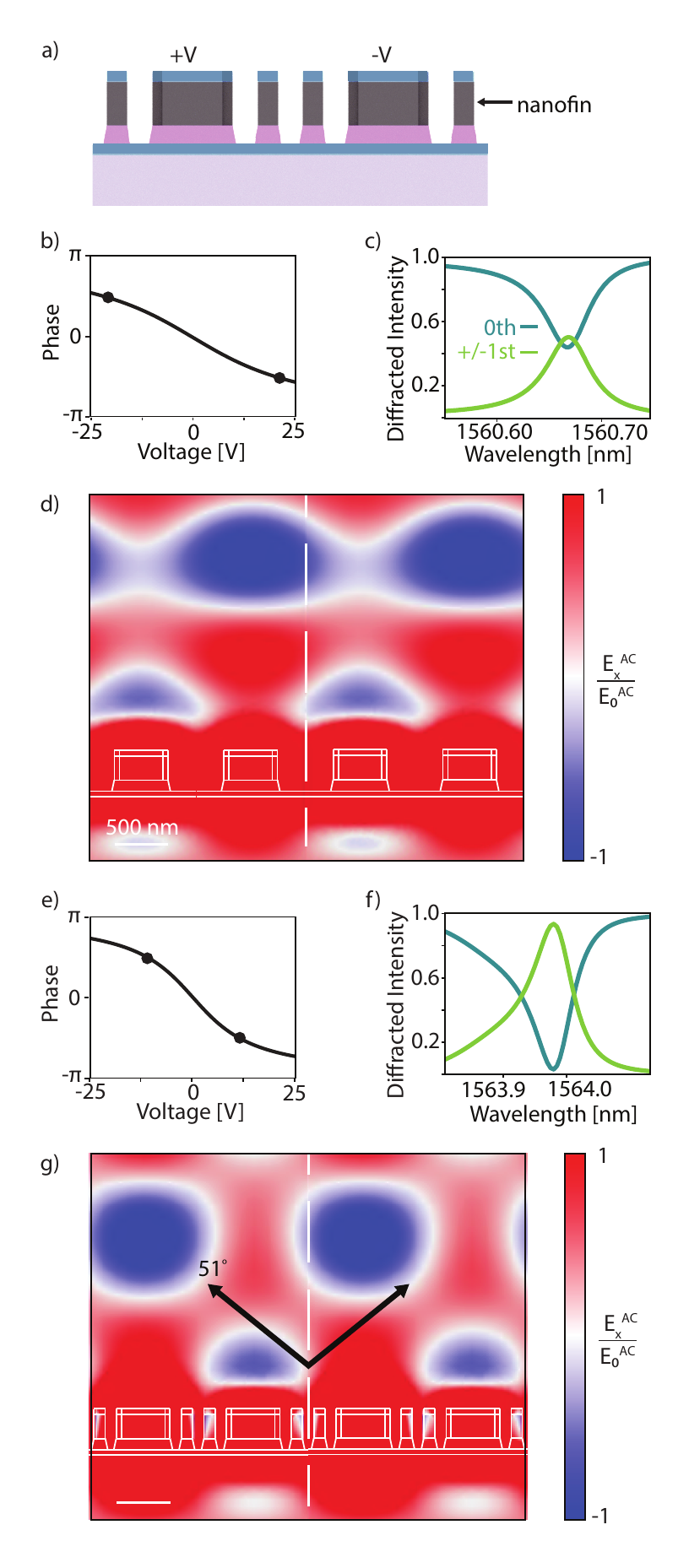}
\caption{Electro-optically switchable metasurface beamsplitter. (a) Schematic applying alternating voltages to individual nanobars. (b, e) Phase delay from applied voltage to the metasurface. (b) Without nanofins ($\lambda$=1560.65 nm), two voltages ($\pm$21 V) are selected (black dots) to achieve a $\pi$ phase change between bars. (e) With nanofins ($\lambda$=1563.96 nm), two voltages ($\pm$11.3 V) are selected (black dots) to achieve a $\pi$ phase change between bars. (c, f) Diffracted intensity at the resonant wavelength for the metasurface (c) without nanofins and (f) with nanofins. (d) Electromagnetic field resulting from metasurface without nanofins operating at $\lambda$=1560.65 nm with alternating applied biases of $\pm$21 V. (g) Electromagnetic field resulting from metasurface with nanofins operating at $\lambda$=1563.96 nm with alternating applied biases of $\pm$11.3 V. All scale bars are 500 nm.}
\end{figure}

Next, we demonstrate how this platform enables beamsteering with dynamically tunable angle control. A beamsteerer is constructed by a linear phase gradient spanning 2$\pi$ over the metasurface supercell. Incident light is reflected at an angle $\theta_r$ determined by\cite{Yu2011e}:
\begin{equation}
    \theta_r=arcsin(\frac{\lambda_0}{n_i p}+sin(\theta_i))
\end{equation}
where, $\lambda_0$ is the incident wavelength, $n_i$ is the refractive index of the incident medium, $p$ is the supercell size, and $\theta_i$ is the incident angle. In our metasurface design, each supercell has an integer $n$ number of nanobars within it. Therefore, we look for applied field configurations that give neighbor-to-neighbor phase variation of:
\begin{equation}
    \Delta\phi=\frac{2\pi}{n}
\end{equation}
Figure 4a schematically illustrates one possible supercell of 5 nanobars each with a different applied voltage. By  changing  the  number  of  bars in the supercell period, and thus the size of the supercell period, we control the beamsteering angle operating at the resonant frequency. 

We show how our metasurface can form a tunable beamsteerer by modifying the biasing period of our device. For example, we can use supercells composed of 3 (Figures 4b-d), 4 (Figures 4e-g), or 5 (Figures 4h-j) biasing supercell periods to dynamically change the steering angle. We choose voltages applied to individual bars within the supercell that introduce the desired linear phase variation for each beamsteering direction, using 31{\textdegree}, 23{\textdegree}, and 18{\textdegree} as example angles (Figures 4b, 4e, and 4h). This corresponds to a difference in phase response between neighboring bars of $\frac{2\pi}{3}$, $\frac{2\pi}{4}$, and $\frac{2\pi}{5}$ for the 3, 4, and 5 bar supercells respectively. Figures 4c, f, and i show the calculated reflection into each potential diffraction order, showing high efficiency at the design wavelength (1563.96 nm). Specifically, we demonstrate beamsteering efficiencies of 76\%, 80\%, and 86\%, respectively, as shown by the preferential diffraction to the +1st order. Our high-Q nanobars, with Q>30,000, allow strong and tunable beamsteering with just a few volts applied to most elements. The required bias voltages can be further reduced by increasing the Q, though we note an increased sensitivity to coupling will eventually force a trade-off between biasing efficiency and resolution. 

\begin{figure}[htbp]
\centering\includegraphics[width=7cm]{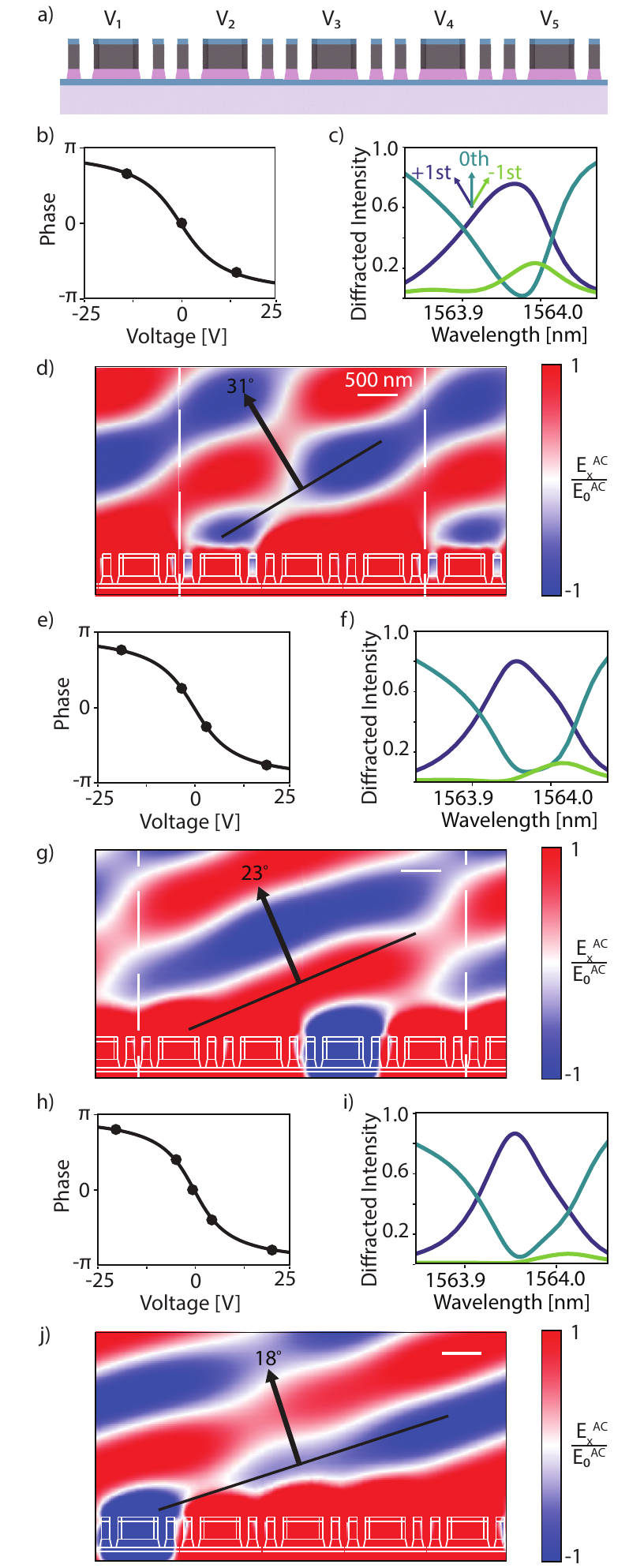}
\caption{Electro-optically reconfigurable metasurface beamsteerer. (a) Schematic applying a series of voltages to individual nanobars for a linear phase gradient across the supercell. (b, e, h) Phase delay from applied voltage. (b) Three voltages ($\pm$14.5, 0 V) are selected (black dots) to achieve a 2$\pi$/3 phase gradient. (e) Four voltages ($\pm$18.6, $\pm$2.7 V) are selected to achieve a 2$\pi$/4 phase gradient. (h) Five voltages ($\pm$21.1, $\pm$5, 0 V) are selected to achieve a 2$\pi$/5 phase gradient. (c, f, i) Diffracted intensity near the resonant wavelength when (c) three, (f) four, and (i) five voltages are applied. (d, g, j) Electromagnetic field resulting from metasurface operating at 1563.96 nm when (d) three, (g) four, and (j) five voltages are applied; white dotted lines denote the supercell period in (d) and (g). All scale bars are 500 nm.}
\end{figure}

We note that this principle can be extended to a broader angle range by increasing the number of bars in the phase gradient, and thus the supercell period. With a single metasurface design, light can be steered to a specific set of angles determined by the period lengths formed by the number of biases in the supercell. To increase the set of angles that can be accessed, the bar width and spacing can be altered (shown in Figure S17). As such, continuous beamsteering could be achievable through a series of diffractive metasurface pixels integrated into a single device. Due to the extended nature of nanobars in one dimension, many phase gradient metasurface designs, including the one presented in this work, use 1D phase gradients to generate the desired transfer function\cite{Aieta2015,Lawrence2020d,Shirmanesh2020d,Park2020e}. The pixelation of metasurfaces also opens the possibility of achieving more complex transfer functions past the limitations of a 1D phase gradient.

Our proposed metasurface utilizes high-Q resonances that are an order of magnitude higher than experimentally reported Q factors in similar structures\cite{Lawrence2020d}. We believe further increasing the Q factor of our resonances is feasible by refining fabrication processes. However, it is known that some losses will necessarily be introduced by fabrication. Figure S18 in the Supplemental Information investigated possible losses from surface roughness, and found that while it may alter the overall efficiencies possible for the device it does not strongly perturb the phase behavior and underlying principles of the device. Furthermore, other electro-optic materials with higher electro-optic coefficients than LNO could also be considered for this design, which would decrease the Q factor needed for efficient modulation. Additionally, implementing design strategies such as array-level inverse design can be used to design devices that can operate within a reduced phase modulation range\cite{Thureja2020}. 

%\section{Conclusion}

In summary, we have demonstrated a metasurface design that individually tunes high-Q resonances in subwavelength nanobars for reconfigurable and highly efficient wavefront shaping. Using full-field coupled simulations, we demonstrate that the electro-optic effect in LNO can be used to control the spectral position of a resonance, and thus its accompanying phase delay. With our Si-on-LNO platform we achieve nearly 2$\pi$ phase variation without sacrificing device efficiency. By modulating the applied electric field profile, we can fully reconfigure our metasurface to act as a beamsplitter or as a beamsteering structure whose steering angle can be dynamically changed with high efficiency (93\% in beamsplitting and 86\% in beamsteering). Moreover, this metasurface design is widely generalizable to other transfer functions, as it can generate any arbitrary phase profile, such as a hyperboloid for future reconfigurable lensing and other non-periodic series. Additionally, a variation of this platform could be explored in transmission through amplitude tuning rather than phase. Our metasurface platform provides a foundation for a multitude of spatial light modulation devices using nanoscale components capable of shaping light as desired.

\begin{acknowledgement}

The authors thank Seagate Technology PLC for their support of this work, especially Zoran Jandric and Aditya Jain for their useful feedback and advice on this manuscript. We gratefully acknowledge support from an AFOSR grant (Grant No. FA9550-20-1-0120), which supported the work and the salaries of D.B., M.L., and J.A.D. E.K. acknowledges support from a National Science Foundation Graduate Research Fellowship Program under grant no. DGE-1656518. S.D. is supported by the Department of Defense (DOD) through the National Defense Science and Engineering (NDSEG) Fellowship Program. 

\end{acknowledgement}

\begin{suppinfo}

Supporting information: details about the guided mode resonance, accounting for coupling, nanofins, pixelation and other design considerations.

\end{suppinfo}

\bibliography{references}

\providecommand{\latin}[1]{#1}
\makeatletter
\providecommand{\doi}
  {\begingroup\let\do\@makeother\dospecials
  \catcode`\{=1 \catcode`\}=2 \doi@aux}
\providecommand{\doi@aux}[1]{\endgroup\texttt{#1}}
\makeatother
\providecommand*\mcitethebibliography{\thebibliography}
\csname @ifundefined\endcsname{endmcitethebibliography}
  {\let\endmcitethebibliography\endthebibliography}{}
\begin{mcitethebibliography}{47}
\providecommand*\natexlab[1]{#1}
\providecommand*\mciteSetBstSublistMode[1]{}
\providecommand*\mciteSetBstMaxWidthForm[2]{}
\providecommand*\mciteBstWouldAddEndPuncttrue
  {\def\EndOfBibitem{\unskip.}}
\providecommand*\mciteBstWouldAddEndPunctfalse
  {\let\EndOfBibitem\relax}
\providecommand*\mciteSetBstMidEndSepPunct[3]{}
\providecommand*\mciteSetBstSublistLabelBeginEnd[3]{}
\providecommand*\EndOfBibitem{}
\mciteSetBstSublistMode{f}
\mciteSetBstMaxWidthForm{subitem}{(\alph{mcitesubitemcount})}
\mciteSetBstSublistLabelBeginEnd
  {\mcitemaxwidthsubitemform\space}
  {\relax}
  {\relax}

\bibitem[Savage(2009)]{Savage2009}
Savage,~N. {Digital spatial light modulators}. \emph{Nature Photonics}
  \textbf{2009}, \emph{3}, 170--172\relax
\mciteBstWouldAddEndPuncttrue
\mciteSetBstMidEndSepPunct{\mcitedefaultmidpunct}
{\mcitedefaultendpunct}{\mcitedefaultseppunct}\relax
\EndOfBibitem
\bibitem[Yu \latin{et~al.}(2011)Yu, Genevet, Kats, Aieta, Tetienne, Capasso,
  and Gaburro]{Yu2011e}
Yu,~N.; Genevet,~P.; Kats,~M.~a.; Aieta,~F.; Tetienne,~J.-P.; Capasso,~F.;
  Gaburro,~Z. {Light Propagation with Phase Reflection and Refraction}.
  \emph{Science} \textbf{2011}, \emph{334}, 333--337\relax
\mciteBstWouldAddEndPuncttrue
\mciteSetBstMidEndSepPunct{\mcitedefaultmidpunct}
{\mcitedefaultendpunct}{\mcitedefaultseppunct}\relax
\EndOfBibitem
\bibitem[Wang \latin{et~al.}(2018)Wang, Kruk, Koshelev, Kravchenko,
  Luther-Davies, and Kivshar]{Wang2018k}
Wang,~L.; Kruk,~S.; Koshelev,~K.; Kravchenko,~I.; Luther-Davies,~B.;
  Kivshar,~Y. {Nonlinear Wavefront Control with All-Dielectric Metasurfaces}.
  \emph{Nano Letters} \textbf{2018}, \emph{18}, 3978--3984\relax
\mciteBstWouldAddEndPuncttrue
\mciteSetBstMidEndSepPunct{\mcitedefaultmidpunct}
{\mcitedefaultendpunct}{\mcitedefaultseppunct}\relax
\EndOfBibitem
\bibitem[Chen \latin{et~al.}(2019)Chen, Zhu, Sisler, Bharwani, and
  Capasso]{Chen2019c}
Chen,~W.~T.; Zhu,~A.~Y.; Sisler,~J.; Bharwani,~Z.; Capasso,~F. {A broadband
  achromatic polarization-insensitive metalens consisting of anisotropic
  nanostructures}. \emph{Nature Communications} \textbf{2019}, \emph{10},
  355\relax
\mciteBstWouldAddEndPuncttrue
\mciteSetBstMidEndSepPunct{\mcitedefaultmidpunct}
{\mcitedefaultendpunct}{\mcitedefaultseppunct}\relax
\EndOfBibitem
\bibitem[Khorasaninejad \latin{et~al.}(2015)Khorasaninejad, Aieta, Kanhaiya,
  Kats, Genevet, Rousso, and Capasso]{Khorasaninejad2015c}
Khorasaninejad,~M.; Aieta,~F.; Kanhaiya,~P.; Kats,~M.~A.; Genevet,~P.;
  Rousso,~D.; Capasso,~F. {Achromatic Metasurface Lens at Telecommunication
  Wavelengths}. \emph{Nano Letters} \textbf{2015}, \emph{15}, 5358--5362\relax
\mciteBstWouldAddEndPuncttrue
\mciteSetBstMidEndSepPunct{\mcitedefaultmidpunct}
{\mcitedefaultendpunct}{\mcitedefaultseppunct}\relax
\EndOfBibitem
\bibitem[Hu \latin{et~al.}(2016)Hu, Liu, Ren, Lauhon, and Odom]{Hu2016}
Hu,~J.; Liu,~C.~H.; Ren,~X.; Lauhon,~L.~J.; Odom,~T.~W. {Plasmonic Lattice
  Lenses for Multiwavelength Achromatic Focusing}. \emph{ACS Nano}
  \textbf{2016}, \emph{10}, 10275--10282\relax
\mciteBstWouldAddEndPuncttrue
\mciteSetBstMidEndSepPunct{\mcitedefaultmidpunct}
{\mcitedefaultendpunct}{\mcitedefaultseppunct}\relax
\EndOfBibitem
\bibitem[Huang \latin{et~al.}(2013)Huang, Chen, M{\"{u}}hlenbernd, Zhang, Chen,
  Bai, Tan, Jin, Cheah, Qiu, Li, Zentgraf, and Zhang]{Huang2013b}
Huang,~L.; Chen,~X.; M{\"{u}}hlenbernd,~H.; Zhang,~H.; Chen,~S.; Bai,~B.;
  Tan,~Q.; Jin,~G.; Cheah,~K.-W.; Qiu,~C.-W.; Li,~J.; Zentgraf,~T.; Zhang,~S.
  {Three-dimensional optical holography using a plasmonic metasurface}.
  \emph{Nature Communications} \textbf{2013}, \emph{4}, 2808\relax
\mciteBstWouldAddEndPuncttrue
\mciteSetBstMidEndSepPunct{\mcitedefaultmidpunct}
{\mcitedefaultendpunct}{\mcitedefaultseppunct}\relax
\EndOfBibitem
\bibitem[Shaltout \latin{et~al.}(2019)Shaltout, Shalaev, and
  Brongersma]{Shaltout2019f}
Shaltout,~A.~M.; Shalaev,~V.~M.; Brongersma,~M.~L. {Spatiotemporal light
  control with active metasurfaces}. \emph{Science} \textbf{2019}, \emph{364},
  eaat3100\relax
\mciteBstWouldAddEndPuncttrue
\mciteSetBstMidEndSepPunct{\mcitedefaultmidpunct}
{\mcitedefaultendpunct}{\mcitedefaultseppunct}\relax
\EndOfBibitem
\bibitem[She \latin{et~al.}(2018)She, Zhang, Shian, Clarke, and
  Capasso]{She2018a}
She,~A.; Zhang,~S.; Shian,~S.; Clarke,~D.~R.; Capasso,~F. {Adaptive metalenses
  with simultaneous electrical control of focal length, astigmatism, and
  shift}. \emph{Science Advances} \textbf{2018}, \emph{4}, eaap9957\relax
\mciteBstWouldAddEndPuncttrue
\mciteSetBstMidEndSepPunct{\mcitedefaultmidpunct}
{\mcitedefaultendpunct}{\mcitedefaultseppunct}\relax
\EndOfBibitem
\bibitem[Wang \latin{et~al.}(2016)Wang, Rogers, Gholipour, Wang, Yuan, Teng,
  and Zheludev]{Wang2016b}
Wang,~Q.; Rogers,~E.~T.; Gholipour,~B.; Wang,~C.~M.; Yuan,~G.; Teng,~J.;
  Zheludev,~N.~I. {Optically reconfigurable metasurfaces and photonic devices
  based on phase change materials}. \emph{Nature Photonics} \textbf{2016},
  \emph{10}, 60--65\relax
\mciteBstWouldAddEndPuncttrue
\mciteSetBstMidEndSepPunct{\mcitedefaultmidpunct}
{\mcitedefaultendpunct}{\mcitedefaultseppunct}\relax
\EndOfBibitem
\bibitem[Howes \latin{et~al.}(2020)Howes, Zhu, Curie, Avila, Wheeler, Haglund,
  and Valentine]{Howes2020b}
Howes,~A.; Zhu,~Z.; Curie,~D.; Avila,~J.~R.; Wheeler,~V.~D.; Haglund,~R.~F.;
  Valentine,~J.~G. {Optical Limiting Based on Huygens' Metasurfaces}.
  \emph{Nano Letters} \textbf{2020}, \emph{20}, 4638--4644\relax
\mciteBstWouldAddEndPuncttrue
\mciteSetBstMidEndSepPunct{\mcitedefaultmidpunct}
{\mcitedefaultendpunct}{\mcitedefaultseppunct}\relax
\EndOfBibitem
\bibitem[Wang \latin{et~al.}(2021)Wang, Landreman, Schoen, Okabe, Marshall,
  Celano, Wong, Park, and Brongersma]{Wang2020a}
Wang,~Y.; Landreman,~P.; Schoen,~D.; Okabe,~K.; Marshall,~A.; Celano,~U.;
  Wong,~H.-S.~P.; Park,~J.; Brongersma,~M.~L. {Electrical tuning of
  phase-change antennas and metasurfaces}. \emph{Nature Nanotechnology}
  \textbf{2021}, \emph{16}, 667--672\relax
\mciteBstWouldAddEndPuncttrue
\mciteSetBstMidEndSepPunct{\mcitedefaultmidpunct}
{\mcitedefaultendpunct}{\mcitedefaultseppunct}\relax
\EndOfBibitem
\bibitem[Shalaginov \latin{et~al.}(2021)Shalaginov, An, Zhang, Yang, Su,
  Liberman, Chou, Roberts, Kang, Rios, Du, Fowler, Agarwal, Richardson,
  Rivero-Baleine, Zhang, Hu, and Gu]{Shalaginov2021a}
Shalaginov,~M.~Y. \latin{et~al.}  {Reconfigurable all-dielectric metalens with
  diffraction-limited performance}. \emph{Nature Communications} \textbf{2021},
  \emph{12}, 1225\relax
\mciteBstWouldAddEndPuncttrue
\mciteSetBstMidEndSepPunct{\mcitedefaultmidpunct}
{\mcitedefaultendpunct}{\mcitedefaultseppunct}\relax
\EndOfBibitem
\bibitem[Zhao \latin{et~al.}(2018)Zhao, Sain, Wei, Tang, Li, Weiss, Huang,
  Wang, and Zentgraf]{Zhao2018}
Zhao,~R.; Sain,~B.; Wei,~Q.; Tang,~C.; Li,~X.; Weiss,~T.; Huang,~L.; Wang,~Y.;
  Zentgraf,~T. {Multichannel vectorial holographic display and encryption}.
  \emph{Light: Science {\&} Applications} \textbf{2018}, \emph{7}, 95\relax
\mciteBstWouldAddEndPuncttrue
\mciteSetBstMidEndSepPunct{\mcitedefaultmidpunct}
{\mcitedefaultendpunct}{\mcitedefaultseppunct}\relax
\EndOfBibitem
\bibitem[Li \latin{et~al.}(2018)Li, Kamin, Zheng, Neubrech, Zhang, and
  Liu]{Li2018}
Li,~J.; Kamin,~S.; Zheng,~G.; Neubrech,~F.; Zhang,~S.; Liu,~N. {Addressable
  metasurfaces for dynamic holography and optical information encryption}.
  \emph{Science Advances} \textbf{2018}, \emph{4}, eaar6768\relax
\mciteBstWouldAddEndPuncttrue
\mciteSetBstMidEndSepPunct{\mcitedefaultmidpunct}
{\mcitedefaultendpunct}{\mcitedefaultseppunct}\relax
\EndOfBibitem
\bibitem[Li \latin{et~al.}(2019)Li, Xu, Veetil, Valuckas,
  Paniagua-Dom{\'{i}}nguez, and Kuznetsov]{Li2019s}
Li,~S.~Q.; Xu,~X.; Veetil,~R.~M.; Valuckas,~V.; Paniagua-Dom{\'{i}}nguez,~R.;
  Kuznetsov,~A.~I. {Phase-only transmissive spatial light modulator based on
  tunable dielectric metasurface}. \emph{Science} \textbf{2019}, \emph{364},
  1087--1090\relax
\mciteBstWouldAddEndPuncttrue
\mciteSetBstMidEndSepPunct{\mcitedefaultmidpunct}
{\mcitedefaultendpunct}{\mcitedefaultseppunct}\relax
\EndOfBibitem
\bibitem[Holsteen \latin{et~al.}(2019)Holsteen, Cihan, and
  Brongersma]{Holsteen2019b}
Holsteen,~A.~L.; Cihan,~A.~F.; Brongersma,~M.~L. {Temporal color mixing and
  dynamic beam shaping with silicon metasurfaces}. \emph{Science}
  \textbf{2019}, \emph{365}, 257--260\relax
\mciteBstWouldAddEndPuncttrue
\mciteSetBstMidEndSepPunct{\mcitedefaultmidpunct}
{\mcitedefaultendpunct}{\mcitedefaultseppunct}\relax
\EndOfBibitem
\bibitem[Wu \latin{et~al.}(2019)Wu, Pala, Kafaie~Shirmanesh, Cheng, Sokhoyan,
  Grajower, Alam, Lee, and Atwater]{Wu2019d}
Wu,~P.~C.; Pala,~R.~A.; Kafaie~Shirmanesh,~G.; Cheng,~W.-H.; Sokhoyan,~R.;
  Grajower,~M.; Alam,~M.~Z.; Lee,~D.; Atwater,~H.~A. {Dynamic beam steering
  with all-dielectric electro-optic III–V multiple-quantum-well
  metasurfaces}. \emph{Nature Communications} \textbf{2019}, \emph{10},
  3654\relax
\mciteBstWouldAddEndPuncttrue
\mciteSetBstMidEndSepPunct{\mcitedefaultmidpunct}
{\mcitedefaultendpunct}{\mcitedefaultseppunct}\relax
\EndOfBibitem
\bibitem[Shirmanesh \latin{et~al.}(2020)Shirmanesh, Sokhoyan, Wu, and
  Atwater]{Shirmanesh2020d}
Shirmanesh,~G.~K.; Sokhoyan,~R.; Wu,~P.~C.; Atwater,~H.~A. {Electro-optically
  Tunable Multifunctional Metasurfaces}. \emph{ACS Nano} \textbf{2020},
  \emph{14}, 6912--6920\relax
\mciteBstWouldAddEndPuncttrue
\mciteSetBstMidEndSepPunct{\mcitedefaultmidpunct}
{\mcitedefaultendpunct}{\mcitedefaultseppunct}\relax
\EndOfBibitem
\bibitem[Park \latin{et~al.}(2021)Park, Jeong, Kim, Lee, Kim, Shin, Lee,
  Otsuka, Kyoung, Kim, Yang, Park, Lee, Hwang, Jang, Song, Brongersma, Ha,
  Hwang, Choo, and Choi]{Park2020e}
Park,~J. \latin{et~al.}  {All-solid-state spatial light modulator with
  independent phase and amplitude control for three-dimensional LiDAR
  applications}. \emph{Nature Nanotechnology} \textbf{2021}, \emph{16},
  69--76\relax
\mciteBstWouldAddEndPuncttrue
\mciteSetBstMidEndSepPunct{\mcitedefaultmidpunct}
{\mcitedefaultendpunct}{\mcitedefaultseppunct}\relax
\EndOfBibitem
\bibitem[Kim \latin{et~al.}(2021)Kim, Martins, Jang, Badloe, Khadir, Jung, Kim,
  Kim, Genevet, and Rho]{Kim2021}
Kim,~I.; Martins,~R.~J.; Jang,~J.; Badloe,~T.; Khadir,~S.; Jung,~H.-Y.;
  Kim,~H.; Kim,~J.; Genevet,~P.; Rho,~J. {Nanophotonics for light detection and
  ranging technology}. \emph{Nature Nanotechnology} \textbf{2021}, \emph{16},
  508--524\relax
\mciteBstWouldAddEndPuncttrue
\mciteSetBstMidEndSepPunct{\mcitedefaultmidpunct}
{\mcitedefaultendpunct}{\mcitedefaultseppunct}\relax
\EndOfBibitem
\bibitem[Huang \latin{et~al.}(2016)Huang, Lee, Sokhoyan, Pala, Thyagarajan,
  Han, Tsai, and Atwater]{Huang2016d}
Huang,~Y.~W.; Lee,~H. W.~H.; Sokhoyan,~R.; Pala,~R.~A.; Thyagarajan,~K.;
  Han,~S.; Tsai,~D.~P.; Atwater,~H.~A. {Gate-Tunable Conducting Oxide
  Metasurfaces}. \emph{Nano Letters} \textbf{2016}, \emph{16}, 5319--5325\relax
\mciteBstWouldAddEndPuncttrue
\mciteSetBstMidEndSepPunct{\mcitedefaultmidpunct}
{\mcitedefaultendpunct}{\mcitedefaultseppunct}\relax
\EndOfBibitem
\bibitem[Jiang \latin{et~al.}(2019)Jiang, Patel, Mayor, McKenna,
  Arrangoiz-Arriola, Sarabalis, Witmer, Van~Laer, and
  Safavi-Naeini]{Jiang2019a}
Jiang,~W.; Patel,~R.~N.; Mayor,~F.~M.; McKenna,~T.~P.; Arrangoiz-Arriola,~P.;
  Sarabalis,~C.~J.; Witmer,~J.~D.; Van~Laer,~R.; Safavi-Naeini,~A.~H. {Lithium
  niobate piezo-optomechanical crystals}. \emph{Optica} \textbf{2019},
  \emph{6}, 845\relax
\mciteBstWouldAddEndPuncttrue
\mciteSetBstMidEndSepPunct{\mcitedefaultmidpunct}
{\mcitedefaultendpunct}{\mcitedefaultseppunct}\relax
\EndOfBibitem
\bibitem[Li \latin{et~al.}(2020)Li, Ling, He, Javid, Xue, and Lin]{Li2020a}
Li,~M.; Ling,~J.; He,~Y.; Javid,~U.~A.; Xue,~S.; Lin,~Q. {Lithium niobate
  photonic-crystal electro-optic modulator}. \emph{Nature Communications}
  \textbf{2020}, \emph{11}, 4123\relax
\mciteBstWouldAddEndPuncttrue
\mciteSetBstMidEndSepPunct{\mcitedefaultmidpunct}
{\mcitedefaultendpunct}{\mcitedefaultseppunct}\relax
\EndOfBibitem
\bibitem[Dabirian \latin{et~al.}(2017)Dabirian, Morales-Masis, Haug, De~Wolf,
  and Ballif]{Dabirian2017}
Dabirian,~A.; Morales-Masis,~M.; Haug,~F.~J.; De~Wolf,~S.; Ballif,~C. {Optical
  Evaluation of the Rear Contacts of Crystalline Silicon Solar Cells by Coupled
  Electromagnetic and Statistical Ray-Optics Modeling}. \emph{IEEE Journal of
  Photovoltaics} \textbf{2017}, \emph{7}, 718--726\relax
\mciteBstWouldAddEndPuncttrue
\mciteSetBstMidEndSepPunct{\mcitedefaultmidpunct}
{\mcitedefaultendpunct}{\mcitedefaultseppunct}\relax
\EndOfBibitem
\bibitem[Jahani and Jacob(2014)Jahani, and Jacob]{Jahani2014a}
Jahani,~S.; Jacob,~Z. {Transparent subdiffraction optics: nanoscale light
  confinement without metal}. \emph{Optica} \textbf{2014}, \emph{1}, 96\relax
\mciteBstWouldAddEndPuncttrue
\mciteSetBstMidEndSepPunct{\mcitedefaultmidpunct}
{\mcitedefaultendpunct}{\mcitedefaultseppunct}\relax
\EndOfBibitem
\bibitem[Jahani \latin{et~al.}(2018)Jahani, Kim, Atkinson, Wirth, Kalhor,
  Noman, Newman, Shekhar, Han, Van, DeCorby, Chrostowski, Qi, and
  Jacob]{Jahani2018b}
Jahani,~S.; Kim,~S.; Atkinson,~J.; Wirth,~J.~C.; Kalhor,~F.; Noman,~A.~A.;
  Newman,~W.~D.; Shekhar,~P.; Han,~K.; Van,~V.; DeCorby,~R.~G.;
  Chrostowski,~L.; Qi,~M.; Jacob,~Z. {Controlling evanescent waves using
  silicon photonic all-dielectric metamaterials for dense integration}.
  \emph{Nature Communications} \textbf{2018}, \emph{9}, 1893\relax
\mciteBstWouldAddEndPuncttrue
\mciteSetBstMidEndSepPunct{\mcitedefaultmidpunct}
{\mcitedefaultendpunct}{\mcitedefaultseppunct}\relax
\EndOfBibitem
\bibitem[Ordal \latin{et~al.}(1987)Ordal, Bell, Alexander, Long, and
  Querry]{Ordal1987}
Ordal,~M.~A.; Bell,~R.~J.; Alexander,~R.~W.; Long,~L.~L.; Querry,~M.~R.
  {Optical properties of Au, Ni, and Pb at submillimeter wavelengths}.
  \emph{Applied Optics} \textbf{1987}, \emph{26}, 744\relax
\mciteBstWouldAddEndPuncttrue
\mciteSetBstMidEndSepPunct{\mcitedefaultmidpunct}
{\mcitedefaultendpunct}{\mcitedefaultseppunct}\relax
\EndOfBibitem
\bibitem[Johnson \latin{et~al.}(1999)Johnson, Fan, Villeneuve, Joannopoulos,
  and Kolodziejski]{Johnson1999c}
Johnson,~S.~G.; Fan,~S.; Villeneuve,~P.~R.; Joannopoulos,~J.~D.;
  Kolodziejski,~L.~A. {Guided modes in photonic crystal slabs}. \emph{Physical
  Review B} \textbf{1999}, \emph{60}, 5751--5758\relax
\mciteBstWouldAddEndPuncttrue
\mciteSetBstMidEndSepPunct{\mcitedefaultmidpunct}
{\mcitedefaultendpunct}{\mcitedefaultseppunct}\relax
\EndOfBibitem
\bibitem[Wang and Magnusson(1993)Wang, and Magnusson]{Wang1993c}
Wang,~S.~S.; Magnusson,~R. {Theory and applications of guided-mode resonance
  filters}. \emph{Applied Optics} \textbf{1993}, \emph{32}, 2606\relax
\mciteBstWouldAddEndPuncttrue
\mciteSetBstMidEndSepPunct{\mcitedefaultmidpunct}
{\mcitedefaultendpunct}{\mcitedefaultseppunct}\relax
\EndOfBibitem
\bibitem[Hsu \latin{et~al.}(2016)Hsu, Zhen, Stone, Joannopoulos, and
  Solja{\v{c}}i{\'{c}}]{Hsu2016d}
Hsu,~C.~W.; Zhen,~B.; Stone,~A.~D.; Joannopoulos,~J.~D.;
  Solja{\v{c}}i{\'{c}},~M. {Bound states in the continuum}. \emph{Nature
  Reviews Materials} \textbf{2016}, \emph{1}, 16048\relax
\mciteBstWouldAddEndPuncttrue
\mciteSetBstMidEndSepPunct{\mcitedefaultmidpunct}
{\mcitedefaultendpunct}{\mcitedefaultseppunct}\relax
\EndOfBibitem
\bibitem[Lawrence \latin{et~al.}(2018)Lawrence, Barton, and
  Dionne]{Lawrence2018i}
Lawrence,~M.; Barton,~D.~R.; Dionne,~J.~A. {Nonreciprocal Flat Optics with
  Silicon Metasurfaces}. \emph{Nano Letters} \textbf{2018}, \emph{18},
  1104--1109\relax
\mciteBstWouldAddEndPuncttrue
\mciteSetBstMidEndSepPunct{\mcitedefaultmidpunct}
{\mcitedefaultendpunct}{\mcitedefaultseppunct}\relax
\EndOfBibitem
\bibitem[Kim \latin{et~al.}(2019)Kim, Kim, and Cahoon]{Kim2019k}
Kim,~S.; Kim,~K.~H.; Cahoon,~J.~F. {Optical Bound States in the Continuum with
  Nanowire Geometric Superlattices}. \emph{Physical Review Letters}
  \textbf{2019}, \emph{122}, 187402\relax
\mciteBstWouldAddEndPuncttrue
\mciteSetBstMidEndSepPunct{\mcitedefaultmidpunct}
{\mcitedefaultendpunct}{\mcitedefaultseppunct}\relax
\EndOfBibitem
\bibitem[Lawrence \latin{et~al.}(2020)Lawrence, Barton, Dixon, Song, van~de
  Groep, Brongersma, and Dionne]{Lawrence2020d}
Lawrence,~M.; Barton,~D.~R.; Dixon,~J.; Song,~J.~H.; van~de Groep,~J.;
  Brongersma,~M.~L.; Dionne,~J.~A. {High quality factor phase gradient
  metasurfaces}. \emph{Nature Nanotechnology} \textbf{2020}, \emph{15},
  956--961\relax
\mciteBstWouldAddEndPuncttrue
\mciteSetBstMidEndSepPunct{\mcitedefaultmidpunct}
{\mcitedefaultendpunct}{\mcitedefaultseppunct}\relax
\EndOfBibitem
\bibitem[Lawrence and Dionne(2019)Lawrence, and Dionne]{Lawrence2019b}
Lawrence,~M.; Dionne,~J.~A. {Nanoscale nonreciprocity via photon-spin-polarized
  stimulated Raman scattering}. \emph{Nature Communications} \textbf{2019},
  \emph{10}, 3297\relax
\mciteBstWouldAddEndPuncttrue
\mciteSetBstMidEndSepPunct{\mcitedefaultmidpunct}
{\mcitedefaultendpunct}{\mcitedefaultseppunct}\relax
\EndOfBibitem
\bibitem[Klopfer \latin{et~al.}(2020)Klopfer, Lawrence, Barton, Dixon, and
  Dionne]{Klopfer2020c}
Klopfer,~E.; Lawrence,~M.; Barton,~D.~R.; Dixon,~J.; Dionne,~J.~A. {Dynamic
  Focusing with High-Quality-Factor Metalenses}. \emph{Nano Letters}
  \textbf{2020}, \emph{20}, 5127--5132\relax
\mciteBstWouldAddEndPuncttrue
\mciteSetBstMidEndSepPunct{\mcitedefaultmidpunct}
{\mcitedefaultendpunct}{\mcitedefaultseppunct}\relax
\EndOfBibitem
\bibitem[Barton \latin{et~al.}(2020)Barton, Hu, Dixon, Klopfer, Dagli,
  Lawrence, and Dionne]{Barton2020c}
Barton,~D.; Hu,~J.; Dixon,~J.; Klopfer,~E.; Dagli,~S.; Lawrence,~M.; Dionne,~J.
  {High-Q nanophotonics: Sculpting wavefronts with slow light}.
  \emph{Nanophotonics} \textbf{2020}, \emph{10}, 83--88\relax
\mciteBstWouldAddEndPuncttrue
\mciteSetBstMidEndSepPunct{\mcitedefaultmidpunct}
{\mcitedefaultendpunct}{\mcitedefaultseppunct}\relax
\EndOfBibitem
\bibitem[Davoyan and Atwater(2020)Davoyan, and Atwater]{Davoyan2020}
Davoyan,~A.; Atwater,~H. {Perimeter-Control Architecture for Optical Phased
  Arrays and Metasurfaces}. \emph{Physical Review Applied} \textbf{2020},
  \emph{14}, 024038\relax
\mciteBstWouldAddEndPuncttrue
\mciteSetBstMidEndSepPunct{\mcitedefaultmidpunct}
{\mcitedefaultendpunct}{\mcitedefaultseppunct}\relax
\EndOfBibitem
\bibitem[Weis and Gaylord(1985)Weis, and Gaylord]{Weis1985c}
Weis,~R.~S.; Gaylord,~T.~K. {Lithium niobate: Summary of physical properties
  and crystal structure}. \emph{Applied Physics A Solids and Surfaces}
  \textbf{1985}, \emph{37}, 191--203\relax
\mciteBstWouldAddEndPuncttrue
\mciteSetBstMidEndSepPunct{\mcitedefaultmidpunct}
{\mcitedefaultendpunct}{\mcitedefaultseppunct}\relax
\EndOfBibitem
\bibitem[Wang \latin{et~al.}(2019)Wang, Zhang, Yu, Zhu, Hu, and
  Loncar]{Wang2019e}
Wang,~C.; Zhang,~M.; Yu,~M.; Zhu,~R.; Hu,~H.; Loncar,~M. {Monolithic lithium
  niobate photonic circuits for Kerr frequency comb generation and modulation}.
  \emph{Nature Communications} \textbf{2019}, \emph{10}, 978\relax
\mciteBstWouldAddEndPuncttrue
\mciteSetBstMidEndSepPunct{\mcitedefaultmidpunct}
{\mcitedefaultendpunct}{\mcitedefaultseppunct}\relax
\EndOfBibitem
\bibitem[Chen \latin{et~al.}(2013)Chen, Wood, and Reano]{Chen2013g}
Chen,~L.; Wood,~M.~G.; Reano,~R.~M. {125 pm/V hybrid silicon and lithium
  niobate optical microring resonator with integrated electrodes}. \emph{Optics
  Express} \textbf{2013}, \emph{21}, 27003\relax
\mciteBstWouldAddEndPuncttrue
\mciteSetBstMidEndSepPunct{\mcitedefaultmidpunct}
{\mcitedefaultendpunct}{\mcitedefaultseppunct}\relax
\EndOfBibitem
\bibitem[Weigel \latin{et~al.}(2018)Weigel, Zhao, Fang, Al-Rubaye, Trotter,
  Hood, Mudrick, Dallo, Pomerene, Starbuck, DeRose, Lentine, Rebeiz, and
  Mookherjea]{Weigel2018d}
Weigel,~P.~O.; Zhao,~J.; Fang,~K.; Al-Rubaye,~H.; Trotter,~D.; Hood,~D.;
  Mudrick,~J.; Dallo,~C.; Pomerene,~A.~T.; Starbuck,~A.~L.; DeRose,~C.~T.;
  Lentine,~A.~L.; Rebeiz,~G.; Mookherjea,~S. {Bonded thin film lithium niobate
  modulator on a silicon photonics platform exceeding 100 GHz 3-dB electrical
  modulation bandwidth}. \emph{Optics Express} \textbf{2018}, \emph{26},
  23728\relax
\mciteBstWouldAddEndPuncttrue
\mciteSetBstMidEndSepPunct{\mcitedefaultmidpunct}
{\mcitedefaultendpunct}{\mcitedefaultseppunct}\relax
\EndOfBibitem
\bibitem[Witmer \latin{et~al.}(2017)Witmer, Valery, Arrangoiz-Arriola,
  Sarabalis, Hill, and Safavi-Naeini]{Witmer2017d}
Witmer,~J.~D.; Valery,~J.~A.; Arrangoiz-Arriola,~P.; Sarabalis,~C.~J.;
  Hill,~J.~T.; Safavi-Naeini,~A.~H. {High-Q photonic resonators and
  electro-optic coupling using silicon-on-lithium-niobate}. \emph{Scientific
  Reports} \textbf{2017}, \emph{7}, 46313\relax
\mciteBstWouldAddEndPuncttrue
\mciteSetBstMidEndSepPunct{\mcitedefaultmidpunct}
{\mcitedefaultendpunct}{\mcitedefaultseppunct}\relax
\EndOfBibitem
\bibitem[Barton \latin{et~al.}(2021)Barton, Lawrence, and Dionne]{Barton2021d}
Barton,~D.; Lawrence,~M.; Dionne,~J. {Wavefront shaping and modulation with
  resonant electro-optic phase gradient metasurfaces}. \emph{Applied Physics
  Letters} \textbf{2021}, \emph{118}, 071104\relax
\mciteBstWouldAddEndPuncttrue
\mciteSetBstMidEndSepPunct{\mcitedefaultmidpunct}
{\mcitedefaultendpunct}{\mcitedefaultseppunct}\relax
\EndOfBibitem
\bibitem[Aieta \latin{et~al.}(2015)Aieta, Kats, Genevet, and
  Capasso]{Aieta2015}
Aieta,~F.; Kats,~M.~A.; Genevet,~P.; Capasso,~F. {Multiwavelength achromatic
  metasurfaces by dispersive phase compensation}. \emph{Science} \textbf{2015},
  \emph{347}, 1342--1345\relax
\mciteBstWouldAddEndPuncttrue
\mciteSetBstMidEndSepPunct{\mcitedefaultmidpunct}
{\mcitedefaultendpunct}{\mcitedefaultseppunct}\relax
\EndOfBibitem
\bibitem[Thureja \latin{et~al.}(2020)Thureja, Shirmanesh, Fountaine, Sokhoyan,
  Grajower, and Atwater]{Thureja2020}
Thureja,~P.; Shirmanesh,~G.~K.; Fountaine,~K.~T.; Sokhoyan,~R.; Grajower,~M.;
  Atwater,~H.~A. {Array-Level Inverse Design of Beam Steering Active
  Metasurfaces}. \emph{ACS Nano} \textbf{2020}, \emph{14}, 15042--15055\relax
\mciteBstWouldAddEndPuncttrue
\mciteSetBstMidEndSepPunct{\mcitedefaultmidpunct}
{\mcitedefaultendpunct}{\mcitedefaultseppunct}\relax
\EndOfBibitem
\end{mcitethebibliography}

\newpage

\section{TOC graphic}
\begin{figure}[htbp]
\centering\includegraphics[width=7cm]{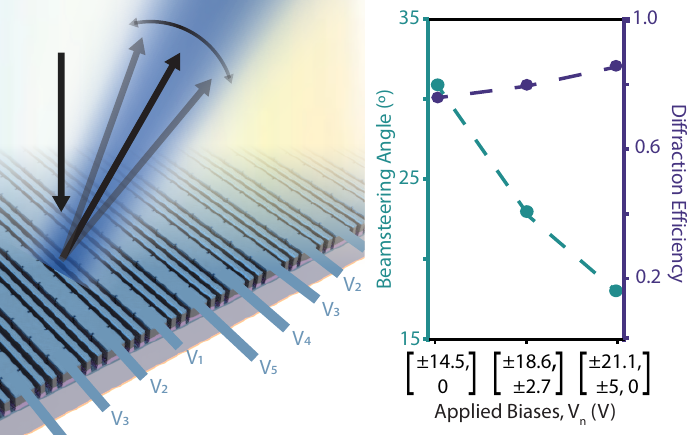}
\end{figure}

\end{document}

\section{Origin of the guided mode resonance}

We leverage high quality factor guided mode resonances to achieve the phase tunability in our metasurface antennas. The Si layer in our metasurface antenna is capable of supporting guided modes, and can be thought of as a waveguide. The calculated dispersion of a Si waveguide is shown in Fig. S1 as the blue curve. Adding periodic perturbations in the form of notches imposes a Bloch condition on the dispersion curve, folding it at the edge of the first Brillouin zone (dashed line) such that it intersects the y-axis at a specific wavelength, shown as the cyan curve. The notches provide free space, normally incident light the momentum required to excite this guided mode resonance at the wavelength indicated by the black star. 

\begin{figure}[htbp]
\centering\includegraphics[width=.6\linewidth]{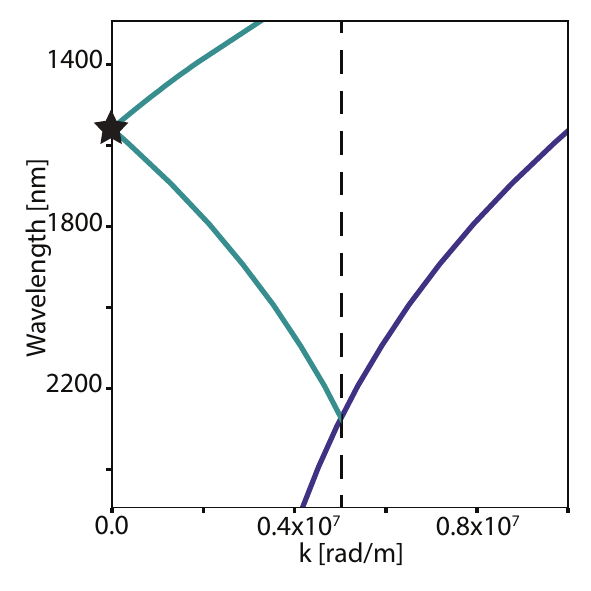}
\caption{Waveguide dispersion of unnotched bar (blue) and notched bar (cyan).}
\end{figure}

\section{Using the reflecting metal plate to expand phase modulation range}

To construct linear phase gradients in our metasurface, a phase modulation range spanning 2$\pi$ is desired. We simulate the transmitted and reflected spectra for a metasurfaces without metal reflecting layer (transmitted) and with reflecting layer (reflected), shown in Fig. 2a and b, respectively. Across the resonance in transmission, the phase response spans a range of $\pi$. Across the resonance in reflection, the phase response spans a range of 2$\pi$. This extra $\pi$ of range in reflection is a result of the $\pi$ phase shift from light reflecting off the metal layer. This expanded phase range allows us to construct phase gradients that span the range required to realize beamsteering and beamsplitting, as demonstrated in the main text, as well as maintains the potential for more complex phase gradients necessary for lensing and beyond.

\begin{figure}[htbp]
\centering\includegraphics[width=\linewidth]{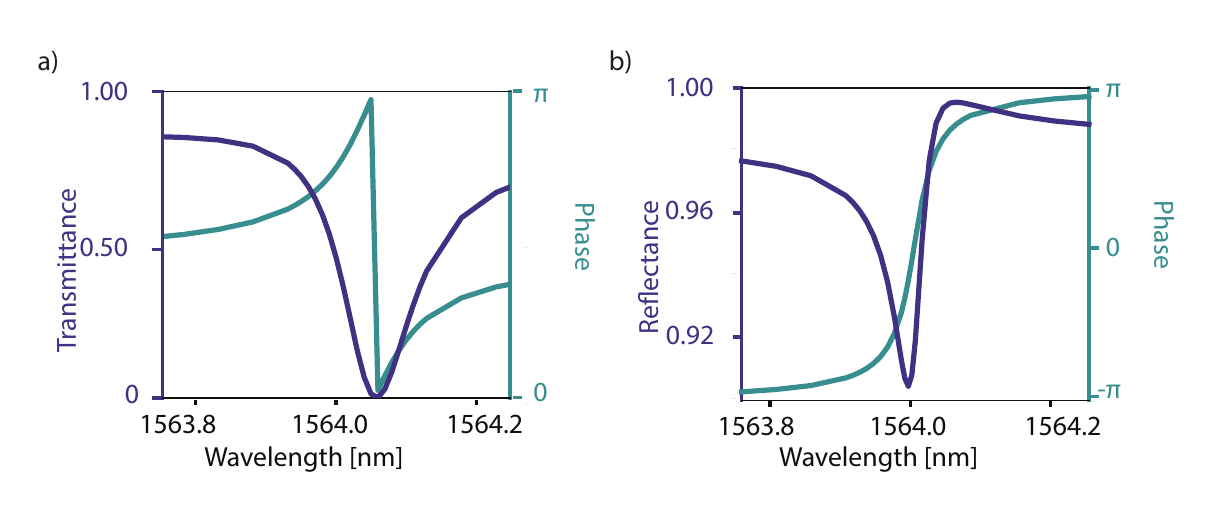}
\caption{(a) Transmittance and phase response across the resonance in a metasurface without a metal reflecting layer and (b) reflectance and phase response across the resonance in a metasurface with a metal reflecting layer separated from the metasurface by a 610 nm SiO$_2$ layer.}
\end{figure}

The Fabry-Perot behavior of light passing through the SiO$_2$ layer affects the guided mode resonance. In Fig. S3, we calculate the absorption across the resonance with varying metal layer depths below the bottom ground transparent conducting oxide contact. We observe that the linewidth of the resonance, and thus the Q factor, and absorption are both variable to the metal plate depth. As such, the SiO$_2$ thickness is optimized to balance achieving a high-Q resonance that can be modulated with a reasonable applied voltage range while keeping absorption low to maximize the diffracted efficiency. 

\begin{figure}[htbp]
\centering\includegraphics[width=.8\linewidth]{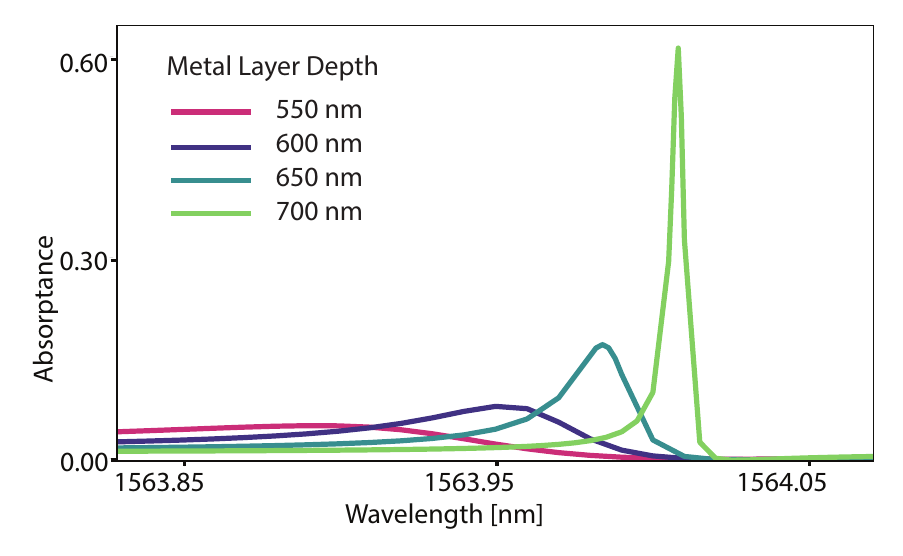}
\caption{Absorptance across the resonance with varying metal plate depths}
\end{figure}

\section{Fins for optical and electronic isolation}

Arraying resonant nanoantennas with subwavelength spacing can result in crosstalk between neighboring antennas and loss of their individual addressability through coupling. To suppress this coupling and ensure the high-Q  modes operate independently  from  one  another while maintaining subwavelength separation, we include  a series of “nanofins” between each nanobar. The  nanofins  are  100  nm  wide  with  a 100  nm center-to-center space between them, so that they act as an effective anisotropic medium to prevent the horizontal high-Q optical modes from coupling to modes in neighboring nanobars. 

Figure S4 plots the near field on resonance with two voltages applied for a metasurface (a) without fins and (b) with fins. The metasurface without fins shows near fields with nearly identical magnitudes in neighboring bars. This is due to the the highly resonant bars couple together when no fins are included, and are therefore not able to be individually tuned. By contrast the metasurface with fins shows resonances with varying magnitude in neighboring bars indicating significant and independent shifts in the resonance due to the applied voltage.  A similar effect is seen in Figure S5, when three voltages are applied (a) without fins vs (b) with fins, as well for the case when (c) four or (d) five voltages are applied. This optical isolation we get from our fins is critical for individual addressability between neighboring high-Q bars while maintaining subwavelength separation.

\begin{figure}[htbp]
\centering\includegraphics[width=.8\linewidth]{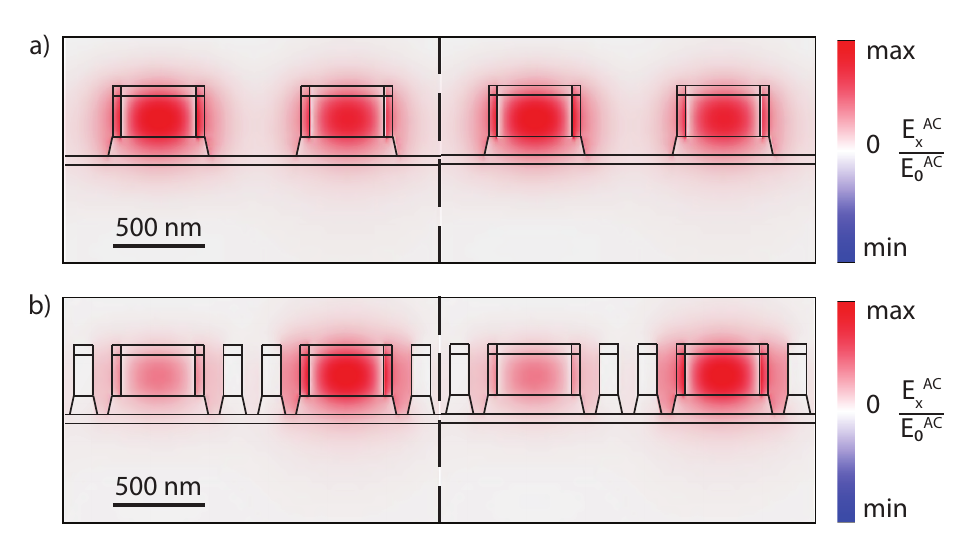}
\caption{Near field plots for a metasurface on resonance with two voltages applied (a) without fins and (b) with fins.}
\end{figure}

\begin{figure}[htbp]
\centering\includegraphics[width=.8\linewidth]{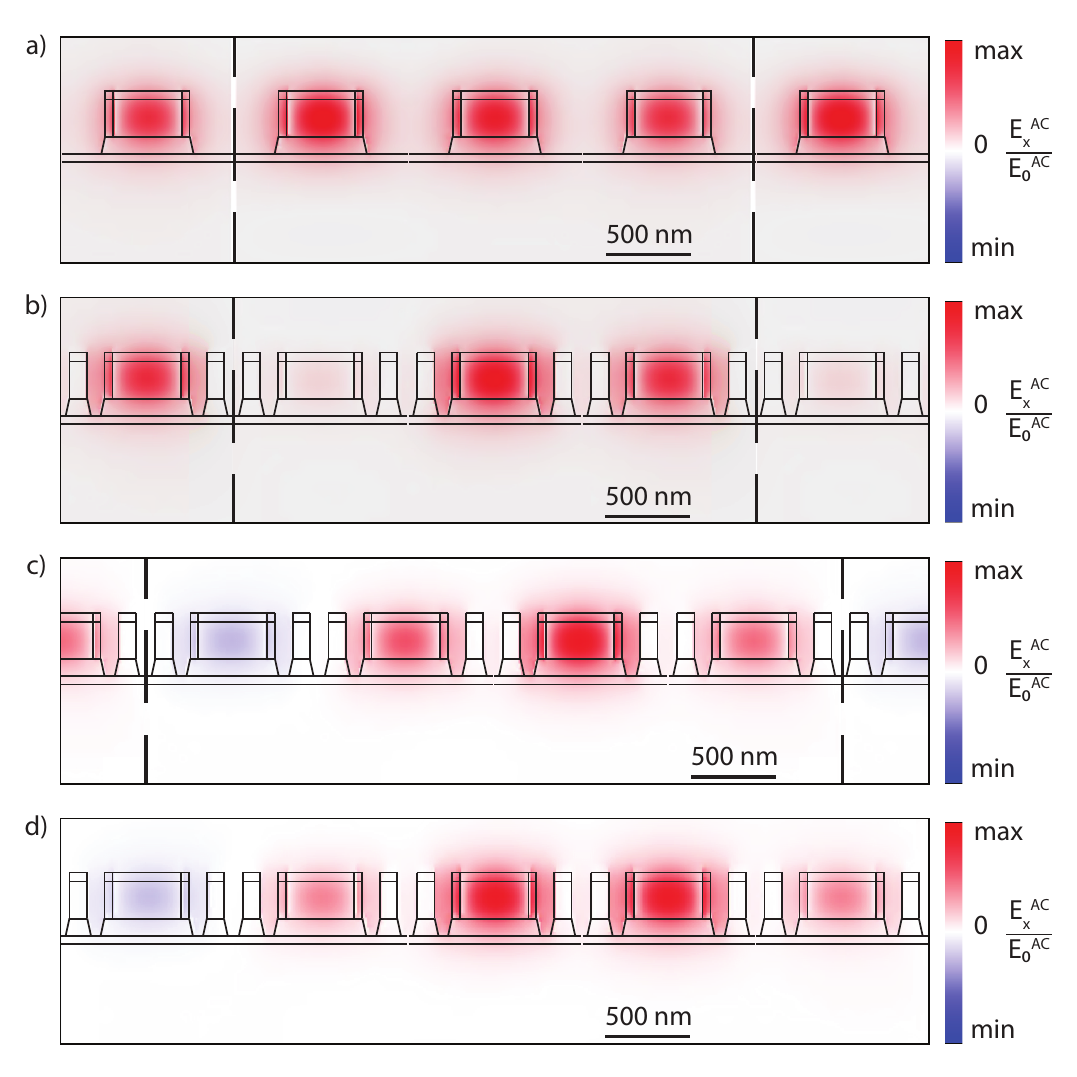}
\caption{Near field plots for a metasurface on resonance (a) with no fins and three voltages applied, and with fins with (b) three, (c) four, and (d) five voltages applied.}
\end{figure}

In addition to the fins improving individual addressability of neighboring antennas by reducing coupling, they also reduce the amount of voltage leaking to neighboring bars significantly. Fig. S6 (a) shows the DC electric field as a function of voltage for one bar, while Fig. S6 (b) shows the DC electric field in a directly adjacent bar in which no voltage is applied as a function of voltage applied to the original bar. When fins are included, the voltage in the first bar results in a much lower DC field in the neighboring bar than in a metasurface without fins. This demonstrates that the fins are crucial for both optical and electronic isolation. 

\begin{figure}[htbp]
\centering\includegraphics[width=\linewidth]{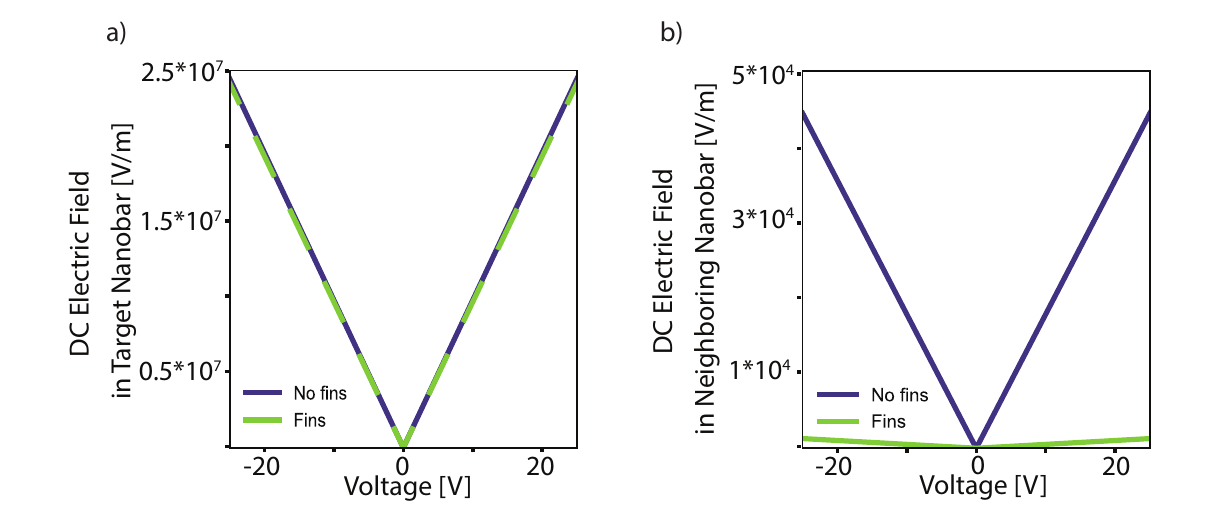}
\caption{Applied DC electric field in (a) a constituent nanobar in the metasurface and (b) the resulting electric field in a neighboring bar with no applied voltage in a metasurface with and without isolating fins}
\end{figure}

\section{Orientation of the lithium niobate (LNO) crystallographic axis}

Lithium niobate (LNO) is a birefringent material, and LNO thin films are commercially available as x-cut, where the optical axis of the crystal is in the plane of the film, and z-cut, where the optical axis is oriented orthogonal to the film surface. In figure S7a and b, we simulate the reflectance with applied voltage at 1563.96 nm and phase response, respectively, to compare the two options. The full resonance is swept through with less applied voltage for x-cut LNO, resulting in a slightly larger phase modulation range with lower voltage. As such, we chose to use x-cut LNO for our design rather than z-cut. 

\begin{figure}[htbp]
\centering\includegraphics[width=\linewidth]{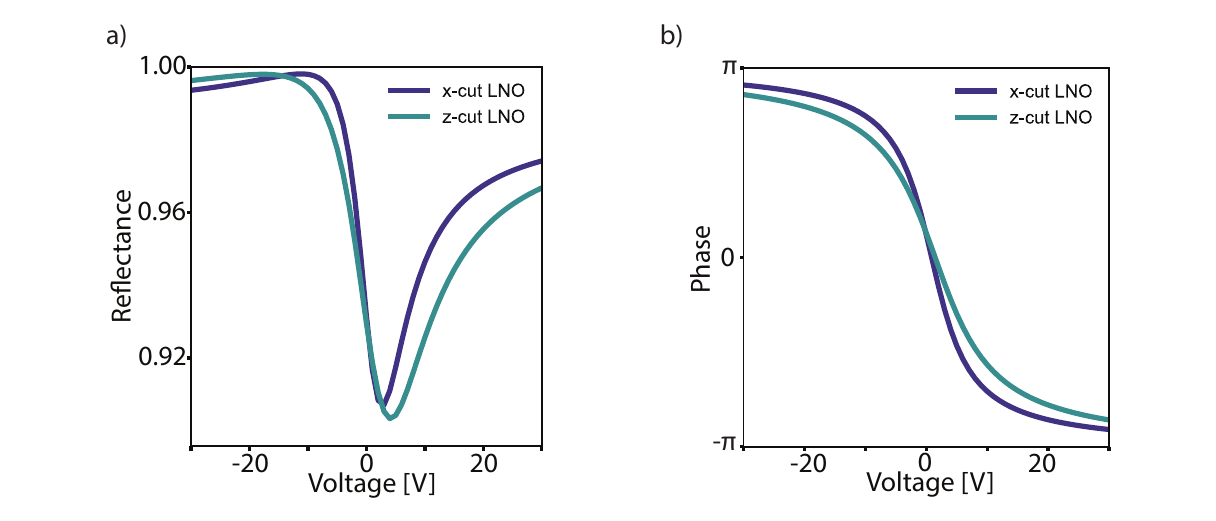}
\caption{(a) Reflectance and (b) phase response across the resonance with metasurfaces using x-cut and z-cut LNO}
\end{figure}

\section{Tunable beamsplitting and beamsteering metasurfaces without accounting for residual coupling}

If we assume that coupling between neighboring bars is negligible, we can use the phase response from one antenna in an infinitely periodic metasurface to design linear phase gradients. We can achieve beamsplitting and beamsteering using this idealized principle, however it results in lower efficiency devices than the platform is otherwise capable of. 

First, we demonstrate beamsplitting by considering a supercell of 2 bars (fig. S8a). Applying voltages of $\pm$ 4.7 V  generates a $\pi$ phase shift between neighboring bars (fig. S8b), which results in a combined $\pm$ 1st diffraction order efficiency of ~40\%. We then show how our metasurface can form a tunable beamsteerer by modifying the biasing and supercell period of our device. For example, we can use supercells composed of 3 (fig. S9a), 4 (fig. S10a), or 5 (fig. S11a) nanoantennas to dynamically change the steering angle. Using equation 1 in the main text, these correspond to beamsteering angles of 31\textdegree, 23\textdegree, and 18\textdegree, respectfully. To do so, we choose voltages applied to individual bars within the supercell that introduce the desired linear phase variation for each beamsteering angle (figs. S9b, S10b, and S11b). This corresponds to a difference in phase response between neighboring antennas of 2$\pi$/3, 2$\pi$/4, and 2$\pi$/5 for the 3, 4, and 5 bar supercells respectively. Figs. S9c, S10c, and S11c show the calculated reflection into each potential diffraction order, showing high efficiency at the design wavelength ($\lambda$=1563.96 nm). Specifically, we demonstrate beamsteering efficiencies of approximately 60\%, 70\%, and 80\%, respectively, as shown by the preferential diffraction to the +1st order. This demonstrates how constructing phase gradients using the idealized phase response variation found via a uniform voltage applied across the metasurface can still result in beamsteering with efficiencies greater than 50\%. 

\begin{figure}[htbp]
\centering\includegraphics[width=.8\linewidth]{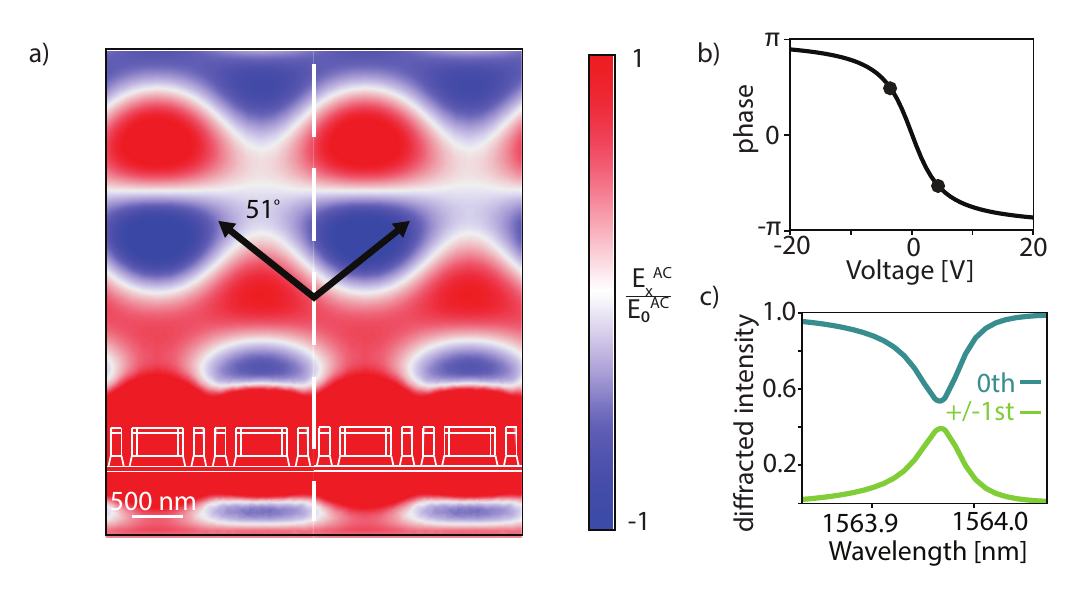}
\caption{Electro-optically reconfigurable metasurface beamsplitter. (a) Visualization of beamsplitting when the series of two voltages are applied to the metasurface; white dotted lines denote the supercell period. (b) Phase delay dictated by applied voltage without accounting for residual coupling between bars, two voltages (-4.7, +4.7 V) are selected (black markers) to achieve a $\pi$ phase change between them. (c) Variation in diffraction efficiency into the $\pm$ 1st diffraction order at and away from the resonant wavelength. On resonance the device operates with a combined $\pm$1st order efficiency of 40\%.}
\end{figure}

\begin{figure}[htbp]
\centering\includegraphics[width=.8\linewidth]{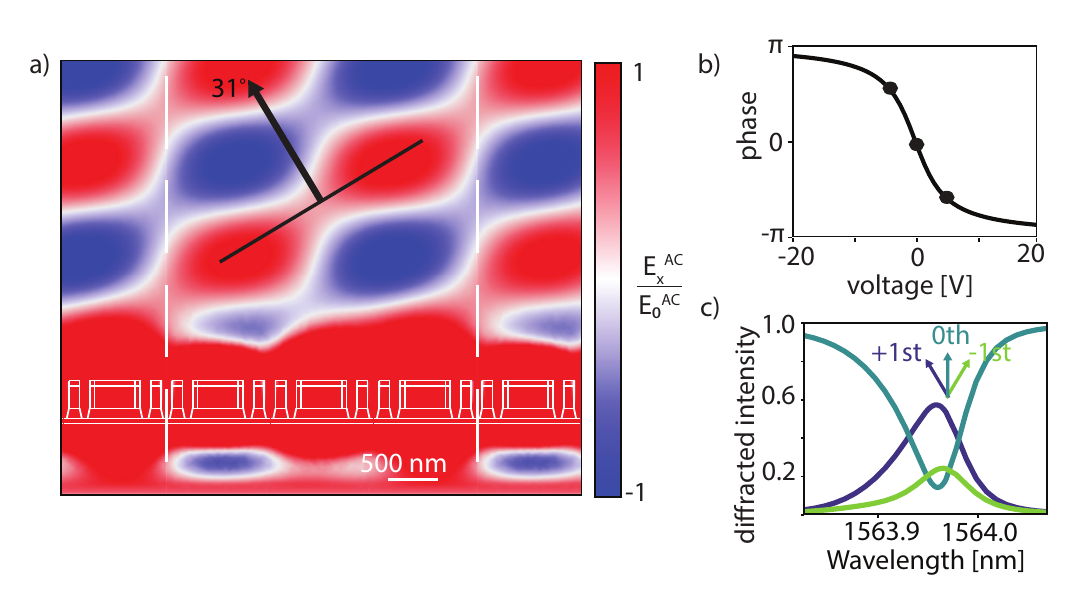}
\caption{Electro-optically reconfigurable metasurface beamsteerer with three bars in the supercell. (a) Visualization of beamsteering at 1563.96 nm when the series of three voltages are applied to the metasurface; white dotted lines denote the supercell period. (b) Phase delay dictated by applied voltage without accounting for residual coupling between bars, three voltages (-8.1, 0, +8.1 V) are selected (black markers) to achieve a 2$\pi$/3 phase change between them. (c) Variation in diffraction efficiency into the +1st diffraction order at and away from the resonant wavelength. On resonance the device operates with a diffracted efficiency of 60\%.}
\end{figure}

\begin{figure}[htbp]
\centering\includegraphics[width=.8\linewidth]{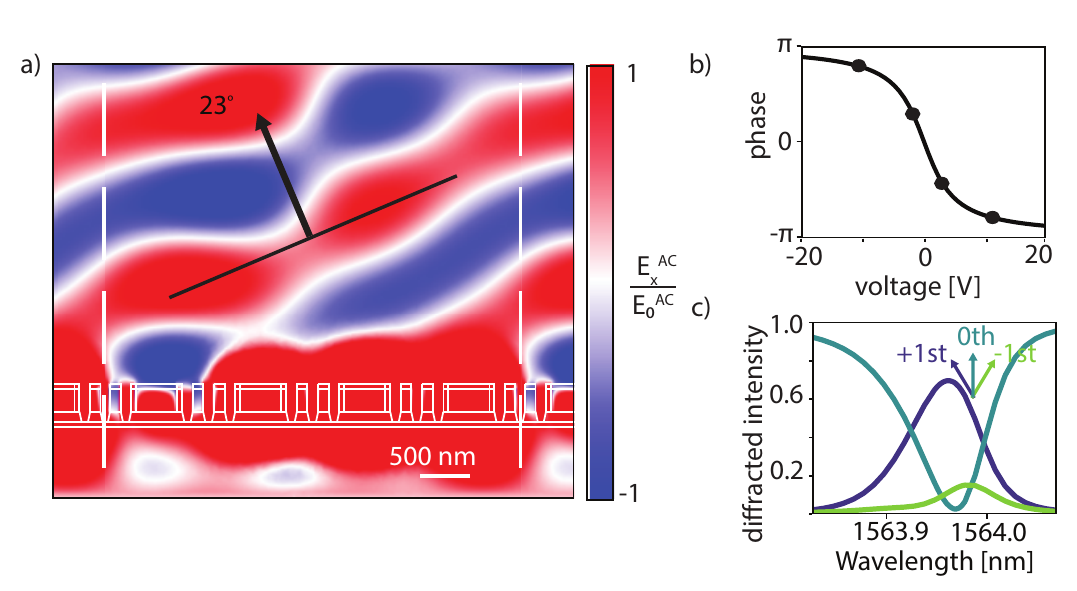}
\caption{Electro-optically reconfigurable metasurface beamsteerer with four bars in the supercell. (a) Visualization of beamsteering at 1563.96 nm when the series of four voltages are applied to the metasurface; white dotted lines denote the supercell period. (b) Phase delay dictated by applied voltage without accounting for residual coupling between bars, four voltages (-11.2, -1.9, +1.9, +11.2 V) are selected (black markers) to achieve a 2$\pi$/4 phase change between them. (c) Variation in diffraction efficiency into the +1st diffraction order at and away from the resonant wavelength. On resonance the device operates with a diffracted efficiency of 70\%.}
\end{figure}

\begin{figure}[htbp]
\centering\includegraphics[width=.8\linewidth]{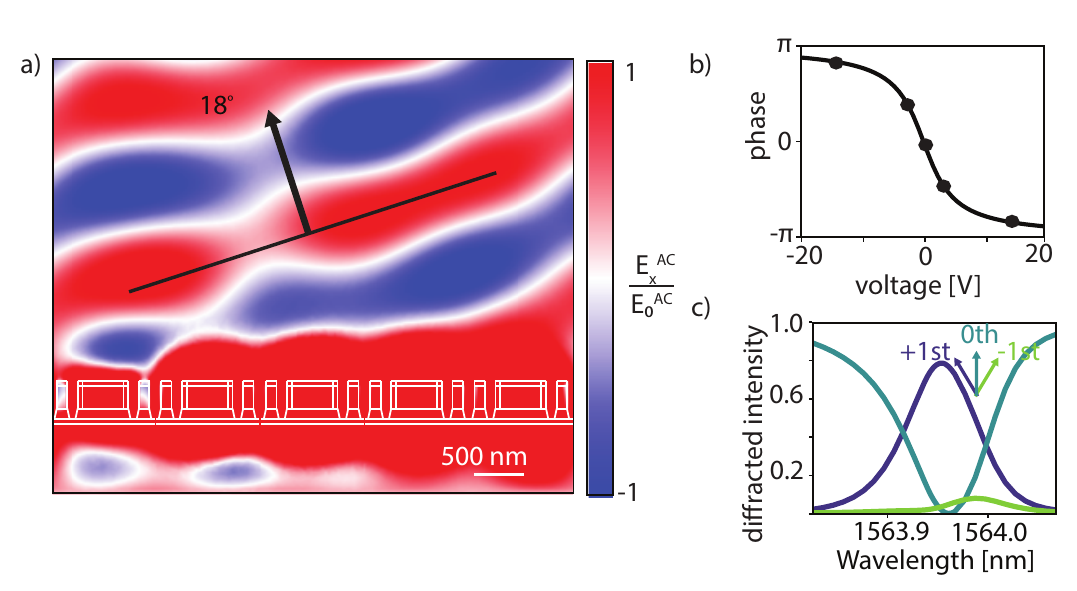}
\caption{Electro-optically reconfigurable metasurface beamsteerer with five bars in the supercell. (a) Visualization of beamsteering at 1563.96 nm when the series of five voltages are applied to the metasurface; white dotted lines denote the supercell period. (b) Phase delay dictated by applied voltage without accounting for residual coupling between bars, five voltages (-14.3, -3.4, 0, +3.4, +14.3 V) are selected (black markers) to achieve a 2$\pi$/5 phase change between them. (c) Variation in diffraction efficiency into the +1st diffraction order at and away from the resonant wavelength. On resonance the device operates with a diffracted efficiency of 80\%.}
\end{figure}

\section{Accounting for residual coupling for highest efficiency beamsplitting and beamsteering}

Although the optically isolating fins allow for individual tuning of neighboring antennas, some coupling will remain in our system. Applying a voltage to perturb resonances in individual bars, this results in a broadening of the resonance. This causes the associated phase response to also broaden with respect to applied voltage. This broadening depends on the phase difference between bars, and we take this coupling into account when calculating the voltages required to tune the phase response of the nanoantennas to achieve the desired transfer function.

When designing our maximum efficiency beamsplitting and beamsteering devices, we account for the coupling between bars that results in the broadening of the phase response from applied voltage by sweeping the applied voltage(s) on the various nanoantennas in the supercell at the design wavelength of 1563.96 nm to maximize the diffracted efficiency to the desired diffraction order. For the beamsplitter formed by applying two biases, this results in a maximized combined $\pm$1st order efficiency of 93\% by applying voltages of $\pm$11.3 V between neighboring bars (fig. S12a). Similarly, for a beamsteerer formed by applying three biases, a maximized +1st diffraction efficiency of 76\% is achieved for applied voltages of $\pm$14.5 V and 0 V (fig S12b). For beamsteerers of larger supercells, multiple sets of voltages must be swept simultaneously. For a beamsteerer formed by applying four biases, a maximized +1st diffraction efficiency of 80\% is achieved for applied voltages of $\pm$18.6 V and $\pm$2.7 (fig S12c). For a beamsteerer formed by applying five biases, a maximized +1st diffraction efficiency of 86\% is achieved for applied voltages of $\pm$21.1 V, $\pm$5, and 0 V (fig S12d). Finally, the curve that defines the relationship between applied voltage and resulting nanoantenna phase response can be adjusted according to the broadening associated with coupling in the case for each supercell (fig S12e).

\begin{figure}[htbp]
\centering\includegraphics[width=.8\linewidth]{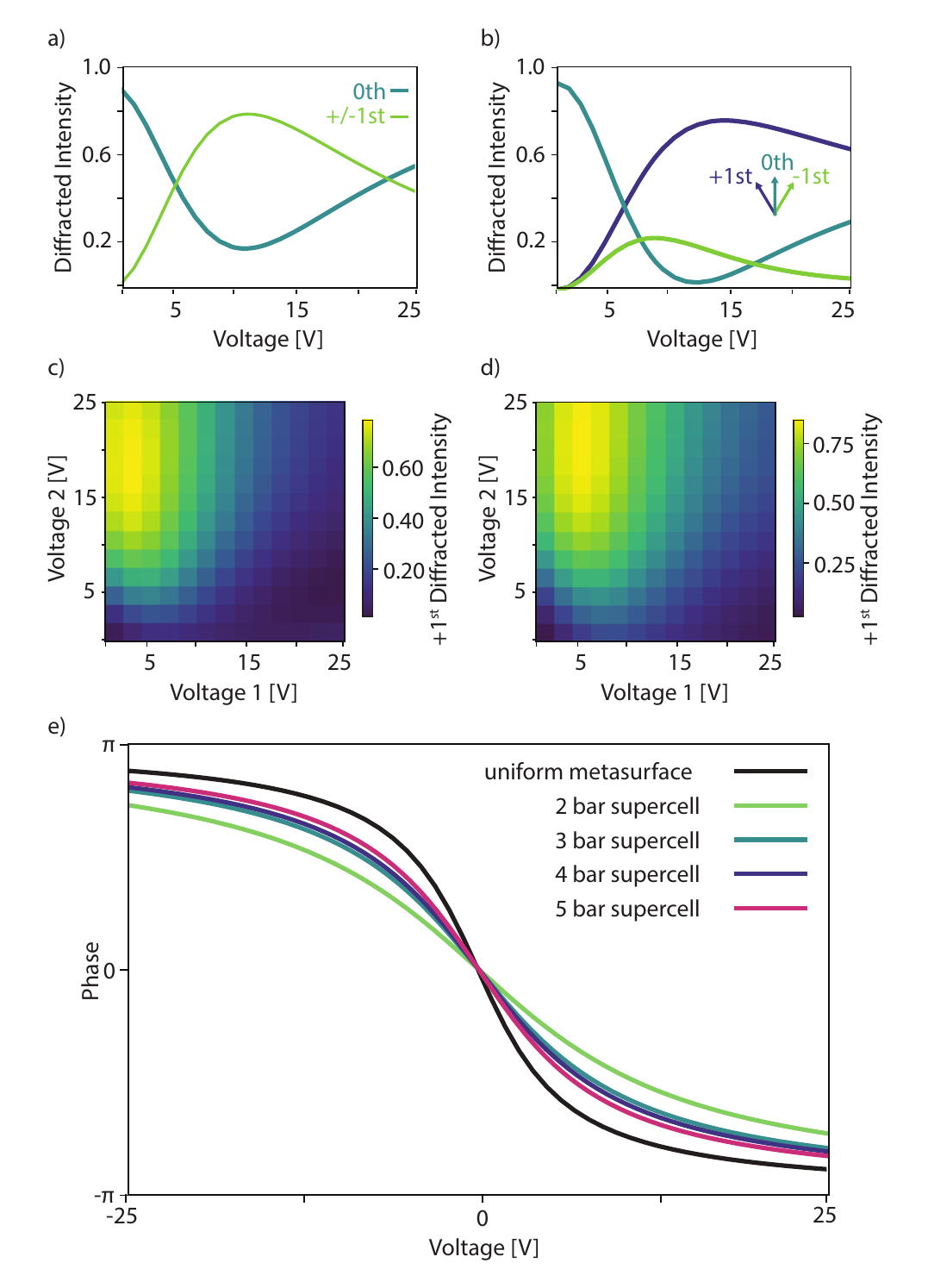}
\caption{Diffracted intensity at 1563.96 nm with respect to voltage applied for a (a) 2 bar supercell with applied voltages of $\pm$V, (b) 3 bar supercell with applied voltages of +V, 0 V, and -V, (c) 4 bar supercell with applied voltages of +Voltage 2, +Voltage 1, -Voltage 1, -Voltage 2, and (d) 5 bar supercell with applied voltages of +Voltage 2, +Voltage 1, 0 V, -Voltage 1, -Voltage 2. (e) Broadening of the phase response with applied voltage modeled for different supercell sizes, resulting from varying amounts of coupling in each case.}
\end{figure}

Interestingly, beamsplitting behavior improves slightly off resonance at 1563.99 nm, achieving a maximum combined $\pm$1st order efficiency of 96\% for applied voltages of $\pm$14.3 V (fig. S13a). This could be accounted for by the increase in overall reflected light from 91\% at 1563.96 nm to 98\% at 1563.99 nm. Figure S13b shows the reflected spectra when $\pm$14.3 V are applied to the metasurface, and figure S13c shows the reflected field at 1563.99 nm with $\pm$14.3 V applied. The beamsteering metasurfaces do not show this behavior.

\begin{figure}[htbp]
\centering\includegraphics[width=.8\linewidth]{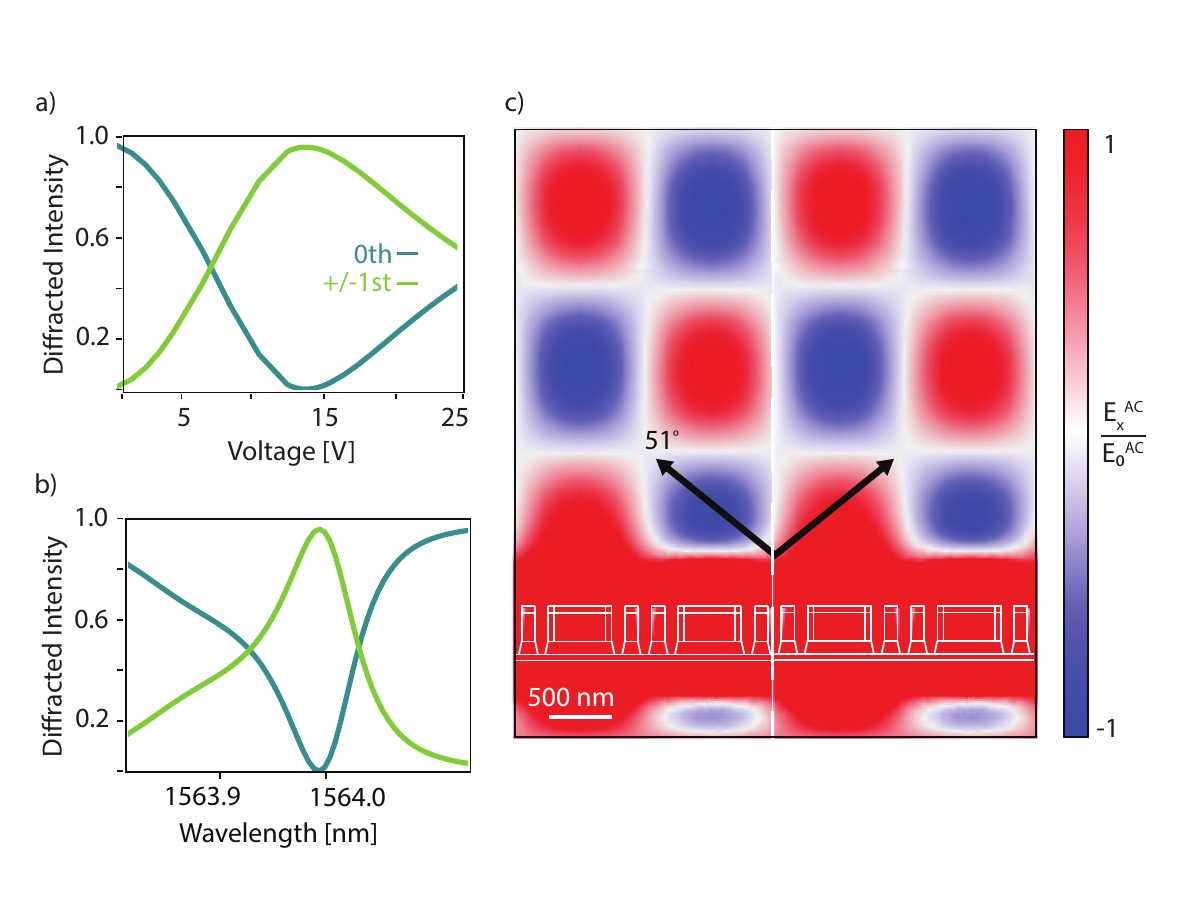}
\caption{(a) Diffracted intensity with respect to applied voltage for a 2 bar supercell at 1563.99 nm. (b) Spectra across the resonance with applied voltages of $\pm$14.3 V. (c) Visualization of beamsteering with the applied voltages at 1563.99 nm.}
\end{figure}

Furthermore, as discussed previously, coupling is exacerbated without the inclusion of nanofins in the system. As such the same applied voltage sweeps are necessary to define the most optimized case for these metasurfaces that experience these coupling effects as well. For the beamsplitter achieved with a 2 bar supercell, this results in a maximized combined $\pm$1st order efficiency of 49\% by applying voltages of $\pm$21 V between neighboring bars (fig. S14a). Similarly, for a beamsteerer achieved with a 3 bar supercell, a maximized +1st diffraction efficiency of 23\% is achieved for applied voltages of $\pm$23 V and 0 V (fig S14b), with 19\% efficiency into the -1st diffraction order. Further details on these metasurfaces, including phase response for the supercell, corresponding reflection in the various diffraction orders across the resonance, and visualization of the resulting field and diffraction angle(s) are found in the main text (fig 3b-d) and below (fig S15).

\begin{figure}[htbp]
\centering\includegraphics[width=\linewidth]{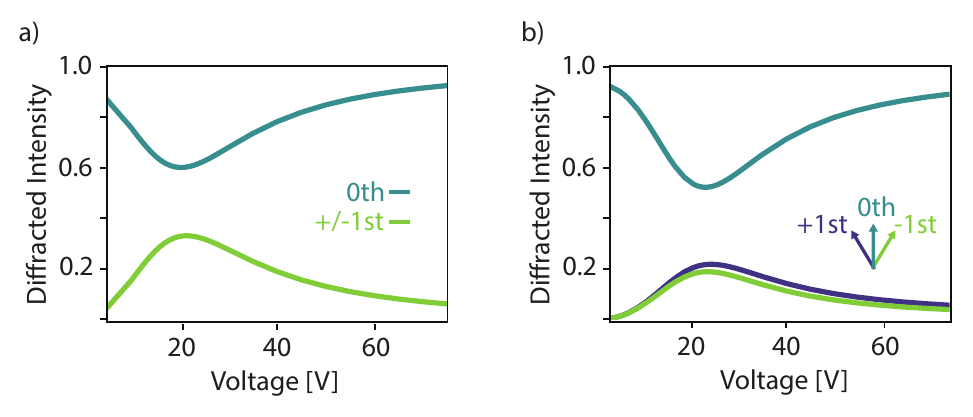}
\caption{Diffracted intensity from a metasurface with no fins between the antennas with respect to applied voltage at 1560.65 nm for a (a) 2 bar supercell and (b) 3 bar supercell.}
\end{figure}

\begin{figure}[htbp]
\centering\includegraphics[width=.8\linewidth]{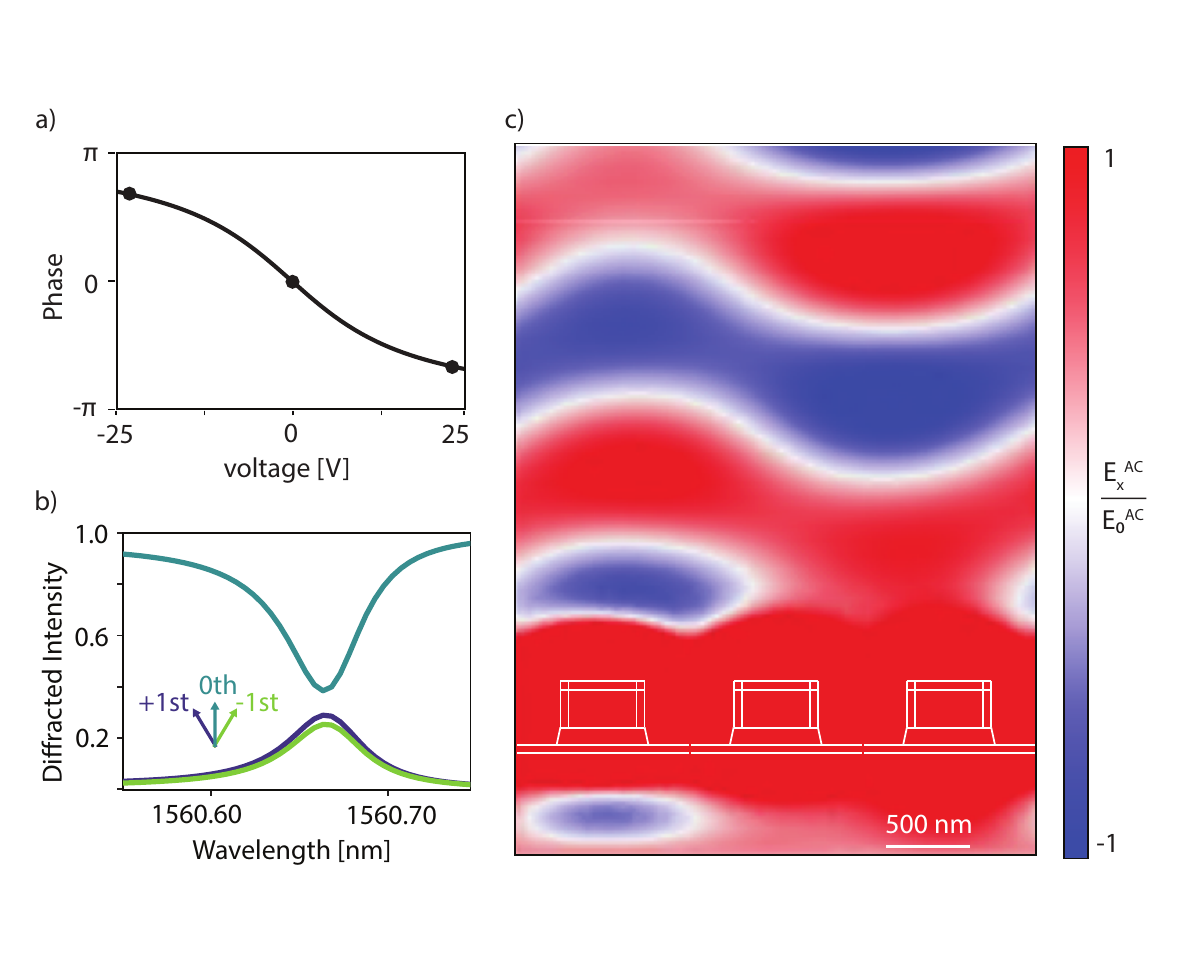}
\caption{(a) Phase delay dictated by applied voltage on a metasurface without fins between the antennas, three voltages (-23 V, 0 V, +23 V) are selected (black markers) to achieve a 2$\pi$/3 phase change between them. (b) Variation in diffraction efficiency at and away from the resonant wavelength. (c) Visualization of the field at 1560.65 nm when the series of three voltages are applied to the metasurface.}
\end{figure}

\section{Extraneous diffraction orders for tunable beamsteerers}

Larger metasurface supercells support diffraction orders beyond the $\pm$1st. When diffraction is maximized in the +1st order at 1563.96 nm, the 4 bar supercell beamsteerer sends less than 5\% of light to the combined $\pm$2nd order diffraction (fig S16a) and the 5 bar supercell beamsteerer sends less than 5\% of light to the $\pm$2nd and 3rd order diffraction (fig S16b). As such, we consider these extraneous diffraction orders negligible in our system.

\begin{figure}[htbp]
\centering\includegraphics[width=\linewidth]{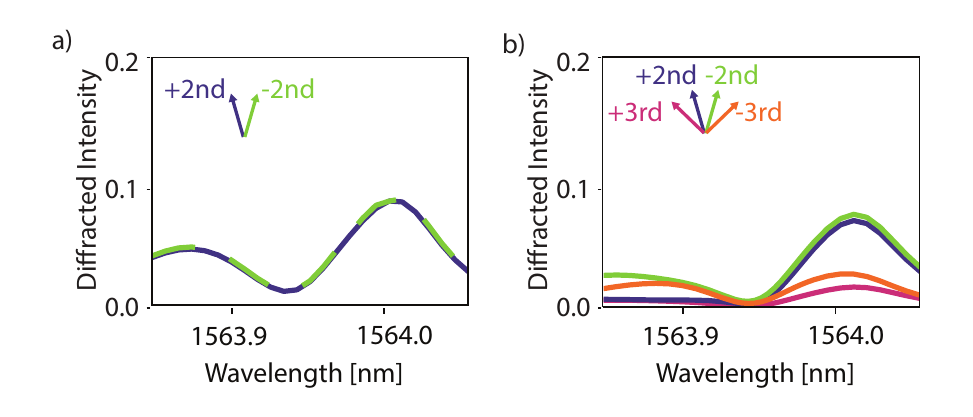}
\caption{(a) Diffraction efficiency into the $\pm$2nd diffraction orders at and away from the resonant wavelength for the tunable beamsteerer reported in Figure 4e-g. (b) Diffraction efficiency into the $\pm$2nd and $\pm$3rd diffraction orders at and away from the resonant wavelength for the tunable beamsteerer reported in Figure 4h-j.}
\end{figure}

\section{Pixelating metasurfaces for continuous angle beamsteering}

Pixelating metasurfaces on a single chip can further enhance the functionality of our tunable metasurface-based devices. In a single metasurface, the number of voltages applied, and thus the number of bars in a supercell period, determine the beamsteering angle. The possible beamsteering angles are determined by the unit cell dimensions of the metasurface. When metasurfaces are pixelated, adjacent pixels can have different unit cell widths resulting from tuning the bar width or spacing to achieve the desired dimensions. This allows each metasurface pixel to access a different set of beamsteering angles, shown in figure S17, enabling near-continuous beamsteering in a single metasurface-based device.

\begin{figure}[htbp]
\centering\includegraphics[width=.6\linewidth]{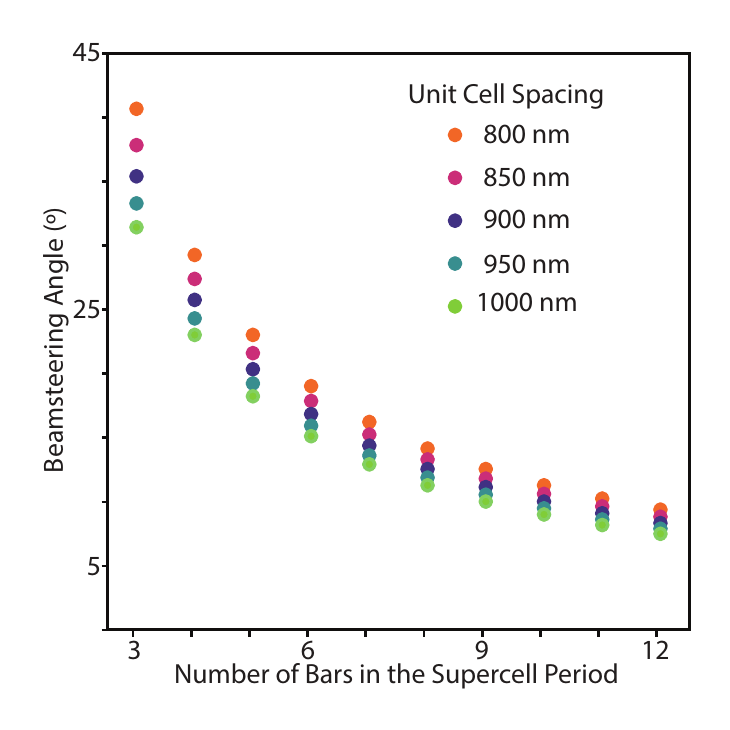}
\caption{Achievable diffractive beamsteering angles based on geometric constraints of the metasurface platform. Inclusion of multiple metasurfaces onto a single device could thus enable more continuous beamsteering.}
\end{figure}

\section{Effect of possible losses introduced by fabrication}

As this work only explores the theoretical design and function of this platform, we generally do not consider the possible losses of the materials involved. While silicon and lithium niobate are dielectrics and do not demonstrate strong losses in the near infrared regime, realizing this platform experimentally will necessarily involve fabrication imperfections and the introductions of losses. For instance, surface roughness can be introduced. Prior work in the literature generally shows high-Qs are feasible in silicon and lithium niobate structures separately\cite{Hang2017,Wang2015,Li2020a}, but surface roughness that may be introduced from fabricating the proposed layered structure may affect the potential Q factor that is able to be achieved in this structure.

Here we consider these possible surface losses by simulating a 10 nm region of the LNO near its bare edges with varying imaginary components of the refractive index (k). As shown in Figure S18b, the introduction of loss affects the reflectance across the resonance strongly. This could impact the maximum potential device efficiency. More importantly, however, the phase variation across the resonance is well preserved at many possible k values. This is likely because the high-Q field is well contained between the Si and LNO, and would not be as strongly perturbed by surface effects.

To move to a fully experimental design,  many features of the current device may need further optimization, as we cannot take into account all possible conditions of fabrication. However, this study supports the core mechanism that underpins our novel approach to dynamic metasurfaces.

\begin{figure}[htbp]
\centering\includegraphics[width=\linewidth]{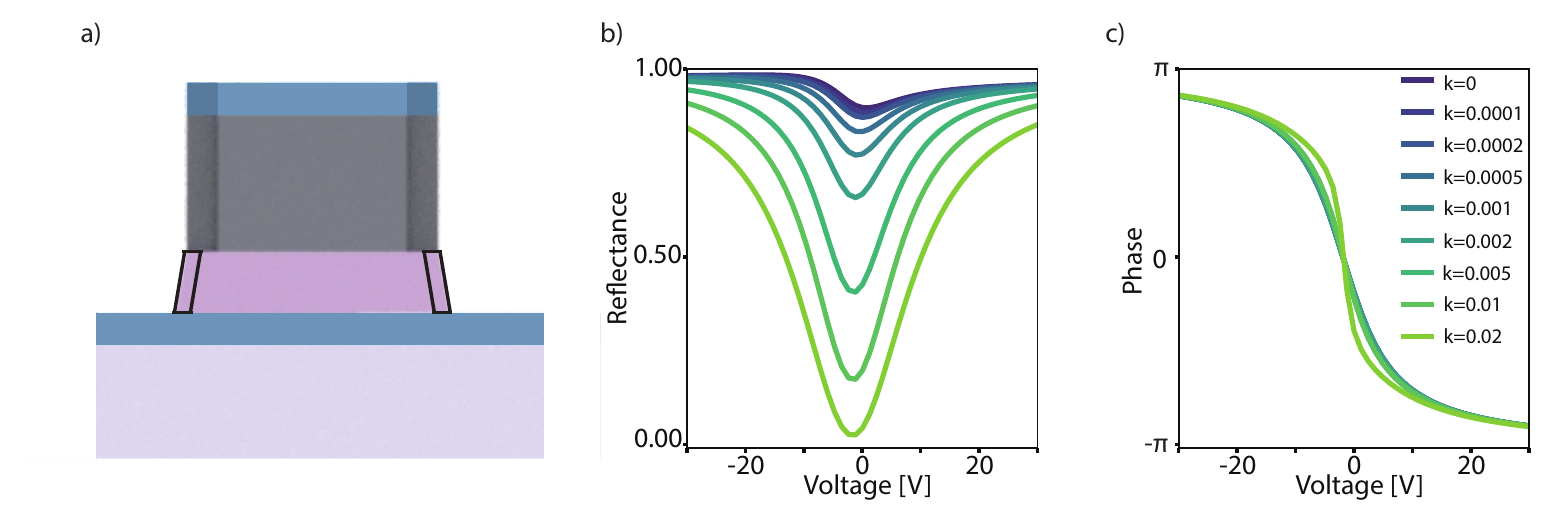}
\caption{ a) Region of surface losses modeled at a 10 nm region b) Spectra of resonant metasurface showing reflectance varying with increasing k (or the imaginary component of the refractive index) c) Phase variation across the resonance for various k values.}
\end{figure}

\clearpage
\bibliography{references}